\documentclass[amsmath, amssymb, aip, jmp, reprint]{revtex4-2}

\usepackage[top=1in, bottom=1.25in, left=1.25in, right=1.25in]{geometry}

\usepackage{tikz}
\usetikzlibrary{shapes.geometric}
\usetikzlibrary{decorations.markings}

\def\tildetriangleleft{\ \widetilde{\triangleleft} \ }

\newtheorem{proposition}{Proposition}
\newtheorem{corollary}{Corollary}

\newcommand{\Lpoly}[1]{%
	\Biggl[
	\begin{tikzpicture}[baseline=-\dimexpr\fontdimen22\textfont2\relax]
		#1
	\end{tikzpicture}%
	\Biggr]_L
}

\newcommand{\Gpoly}[1]{%
	\Biggl[
	\begin{tikzpicture}[baseline=-\dimexpr\fontdimen22\textfont2\relax]
		#1
	\end{tikzpicture}%
	\Biggr]_G
}

\newcommand{\impropervertex}{%
	\Gpoly{%
		\draw[> = latex, ->, dotted, rounded corners] (-0.6, 0.1) -- (-0.2, 0.1) -- (-0.2, 0.6);
		\draw[> = latex, ->, dotted, rounded corners] (0.1, -0.6) -- (0.1, -0.2) -- (0.6, -0.2);

		\draw[> = latex, ->] (0, -0.6) -- (0, 0.6);
		\draw[fill = white] (0, 0) circle (0.15);
		\draw[> = latex, ->] (-0.6, 0) -- (0.6, 0);
	}%
}

\newcommand{\propervertex}{%
	\Gpoly{%
		\draw[> = latex, ->] (0, -0.6) -- (0, 0.6);
		\draw[fill = white] (0, 0) circle (0.15);
		\draw[> = latex, ->] (-0.6, 0) -- (0.6, 0);
	}%
}

\newcommand{\propercross}{%
	\Lpoly{%
		\draw[> = latex, ->] (0, -0.6) -- (0, 0.6);
		\draw[draw = white, double = black, double distance between line centers = 3 pt, line width = 2.6 pt] (-0.6, 0) -- (0.2, 0);
		\draw[> = latex, ->] (0.2, 0) -- (0.6, 0);
	}%
}

\newcommand{\propersplit}{%
	\Lpoly{%
		\draw[> = latex, rounded corners, ->] (-0.6, 0) -- (0, 0) -- (0, 0.6);
		\draw[> = latex, rounded corners, ->] (0, -0.6) -- (0, 0) -- (0.6, 0);
	}%
}

\newcommand{\rvertex}{%
	\Gpoly{%
		\draw[> = latex, ->, dotted, rounded corners] (-0.6, 0.1) -- (-0.2, 0.1) -- (-0.2, 0.6);
		\draw[> = latex, ->, dotted, rounded corners] (0.6, -0.1) -- (0.2, -0.1) -- (0.2, -0.6);

		\draw[> = latex, <->] (0, -0.6) -- (0, 0.6);
		\draw (-0.6, 0) -- (0.6, 0);
		\draw[fill = gray!50] (0, 0) circle (0.15);
	}%
}

\newcommand{\nesplit}{%
	\Lpoly{%
		\draw[> = latex, rounded corners, ->] (-0.6, 0) -- (0, 0) -- (0, 0.6);
		\draw[> = latex, rounded corners, ->] (0.6, 0) -- (0, 0) -- (0, -0.6);
	}%
}

\newcommand{\nwsplit}{%
	\Lpoly{%
		\draw[> = latex, rounded corners, ->] (-0.6, 0) -- (0, 0) -- (0, -0.6);
		\draw[> = latex, rounded corners, ->] (0.6, 0) -- (0, 0) -- (0, 0.6);
	}%
}

\begin{document}

\title{Knotted 4-regular graphs: polynomial invariants and the Pachner moves}
\author{Daniel Cartin}
\email{cartin@naps.edu}

\affiliation{Naval Academy Preparatory School\\440 Meyerkord Ave, Newport, RI, 02841, USA}

\begin{abstract}

In loop quantum gravity, states of quantum geometry are represented by classes of knotted graphs, equivalent under diffeomorphisms. Thus, it is worthwhile to enumerate and distinguish these classes. This paper looks at the case of 4-regular graphs, which have an interpretation as objects dual to triangulations of three-dimensional manifolds. Two different polynomial invariants are developed to characterize these graphs -- one inspired by the Kauffman bracket relations, and the other based on quandles. How the latter invariant changes under the Pachner moves acting on the graphs is then studied.

\end{abstract}

\maketitle





\section{Introduction}
\label{intro}

In loop quantum gravity (LQG), the states of the theory are given by spin networks. These are graphs embedded into a three-dimensional manifold (such as $\mathbb{R}^3$ or $S^3$), with each edge colored with a representation of SU(2). The graphs are known as ``knotted" or ``spatial" graphs, to distinguish their study from abstract graphs, where only the connectivity information of the graph edges is important. In this paper, we will focus on the graph aspect of spin networks. Because these graphs are embedded in a manifold, their placement under this mapping leads to knotting, in the same sense that embedding the circle into a three-dimensional manifold leads to distinct knots. For example, two such graphs $G_1$ and $G_2$ may be isomorphic as abstract graphs, but there is no diffeomorphism of the manifold to itself that takes the embedding of $G_1$ to $G_2$, since it would require edges to pass through each other. Thus, it is worthwhile to examine ways to classify these graphs, in order to understand the space of possible spin networks in LQG.

We will focus in particular on knotted 4-valent graphs, i.e. graphs where each vertex has four edges incident to it. The reasons for choosing valence four graphs will be explained shortly. Note that self-loops and multiple edges are allowed. In other words, the same edge may have both of its ends incident to the same vertex, and two vertices may share more than one edge incident to both. Thus, properly speaking this paper considers pseudographs. However, with this understanding, we will refer to all cases examined as graphs.

In addition, spin networks are properly equivalence classes of graphs, where all members of the class are diffeomorphic to each other. Generically, a consequence of this is that equivalence classes of knotted graphs under diffeomorphisms are labeled by continuous parameters at the vertices~\cite{Gro-Rov96}. Consider the tangent vectors at a given vertex along each of the edges incident to that vertex. The continuous mappings used in knot theory allow these vectors to be moved around arbitrarily, but this is a freedom not present in the smooth map of a diffeomorphism. Thus, there are many possible intersecting links equivalence under diffeomorphisms, for the same equivalence class under continuous maps. In fact, for a large enough valence of the vertex, there are an infinite number of such diffeomorphism-equivalent intersecting links for each ambient isotropic equivalence class of intersecting link. However, this is only true when for vertices with valence five or greater. Thus, choosing all vertices to have valence four alleviates this issue -- the space of graphs with 4-valent vertices, where the edge tangents at the vertices are linearly independent, is countable. Related to this is another consideration arising from LQG. There, the volume operator gives a non-trivial result only when acting on vertices of valence four or higher~\cite{DeP-Rov96}. It should be mentioned that using the space of mappings that are smooth at all but a finite set of points eliminates the problem of linearly dependent tangent vectors for edges as well~\cite{Fai-Rov04}, but we will not use that group extension in this paper.

More physical reasons for choosing 4-valent graphs involve their physical interpretation. First, an abstract 4-valent graph is dual to the simplicial complex of a three-dimensional manifold. This can be easily seen by associating each vertex of the graph with a 3-simplex, and each edge with a face of the simplex. Thus, this allows us to naturally think of the graphs studied in this paper as discrete versions of three-dimensional spaces. In fact, this leads later to the study of graph transformations, modeled on the Pachner moves~\cite{Pac91} acting on simplices. To provide additional context for the Pachner moves, any two triangulations of the same manifold can be related to each other by using a finite sequence of Pachner moves. Thus, in topology they act in a manner similar to the Reidemeister moves of knot theory. The Pachner moves induce a corresponding set of moves on the graphs dual to the triangulation. In this context, they provide a natural set of evolution moves for the graphs used in various models of quantum gravity, such as spin foams~\cite{Dit-Ste14, Ste20}. However, the same abstract graph may be embedded into a manifold in different ways, so there are many such graphs that correspond to the same simplicial complex.

We now provide an outline of this paper. Section \ref{knotted-graphs} goes through the mathematical definitions used through the rest of this work. The graph equivalence moves, generalizing the Reidemeister moves of knot theory, are listed as well. A system of nomenclature for specifying graphs (based on the Dowker-Thistletwaite notation~\cite{Dow-Thi83} from knot theory) is defined in Section \ref{notation}, and it is shown that every graph diagram is equivalent to a diagram where this notation is well-defined. Some physical insight into how the vertices of the graph correspond with the dual triangulation picture will be described. Several simple graph invariants are also defined. Next, two methods for defining polynomial invariants of knotted 4-valent graphs are provided. In Section \ref{skein-poly}, an invariant polynomial for an oriented knotted 4-valent graph is defined, in a manner similar to the Kauffman bracket for the Jones polynomial. This graph polynomial can be thought of as an enhancement of any knot invariant, using two additional abstract variables. Considering all possible Eulerian circuits on a graph allows us to associate an invariant multiset of such graph polynomials to each unoriented graph. A second graph polynomial is defined in Section \ref{quandle-poly} by coloring the arcs of the graph with elements of a quandle. For each Eulerian circuit of the graph, the number of quandle colorings can be computed. The invariant polynomial is then obtained as the generating function for the multiset of such coloring numbers for all possible Eulerian circuits of the graph. Section \ref{Pachner} considers graph transformations based on the Pachner moves on the triangulation dual to the graph. The action of these moves on the quandle invariant polynomial of Section \ref{quandle-poly}, in particular, is studied. Finally, the implications of this paper and some avenues for further research are discussed in Section \ref{discuss}, and the main results are summarized in Section \ref{conclude}.


\section{Knotted graphs}
\label{knotted-graphs}

\subsection{Definitions}
\label{graph-def}

Since many of the terms in graph theory have differing definitions, depending on the reference, here we spell out specifically what is meant by particular terms. Suppose that $G = G(E, V)$ is an abstract graph with a set of vertices $V$, and a set of edges $E$ between the vertices. Thus, labelling each vertex with an index $i$, an unoriented edge is a set of two indices $i, j$, corresponding to the vertices the edge is incident on. If the edges of the graph are oriented, then every edge is considered to be an ordered pair $(i, j)$, where the orientation points from vertex $i$ to vertex $j$. The graphs considered here may have multiple edges between the same two vertices, or have self-loops, where the edge connects twice to the same vertex. These types of graphs are known as {\it pseudographs} (or sometimes {\it multigraphs}). Since all $G$ considered in this work are of this type, from now on they will be referred to as graphs. It is convenient to think of each edge as made of two half-edges, each of which is incident to a vertex and another half-edge. The number of half-edges incident to a particular vertex is known as its {\it valence} or {\it degree}. Then, graphs where all vertices have the same valence are called {\it regular}. All graphs in this work will be {\it 4-regular}, meaning that every vertex has four half-edges incident to it; self-loops thus count as two half-edges incident to the vertex. Because the graphs considered here all have vertices of even degree, there is a well-known theorem that orientations for the graph edges can be chosen so that there is a path traversing the entire graph, passing over each edge (in the direction of its orientation) only once. This is known as an {\it Eulerian circuit} or {\it cycle}.

Let $f: G \to M$ be an embedding of the abstract graph $G$ into a three-dimensional manifold $M$; we will assume that $M$ is $\mathbb{R}^3$ or $S^3$, and not worry about the details of the choice. The image $f(G)$ is referred to as a {\it knotted graph} or {\it spatial graph}. By choosing a plane in $M$, the graph embedding can be projected onto the plane to give a {\it graph diagram} $D$. This will be distinguished in Section \ref{vertex} from an ``induced knot diagram" associated with a graph with an orientation coming from an Eulerian circuit; when the term ``diagram" is used without specification, it is referrring to the graph diagram. There are two reasons that edges in the graph diagram may intersect. One of these is the location of a vertex, the other a {\it crossing}, where one edge passes over the other. The term {\it node} is used to refer to a member of the set of vertices and crossings of the diagram. An {\it arc} is an edge or part of one that connects either two vertices, a vertex and a crossing, or two crossings, so all arcs connect two nodes.

\begin{figure}[hbt]
\centering
\begin{tikzpicture}[> = latex, font = \scriptsize]
\matrix[row sep = 1 cm]{



	\begin{scope}[xshift = -1.5 cm]

		\draw [variable = \t, samples = 100, domain = -90 : 90] plot ({0.5 * sin(\t) - 0.5}, {cos(\t) + 0.3 * (3.14 / 180) * \t});
		\draw [variable = \t, samples = 100, domain = 90 : 270, draw = white, double = black, double distance between line centers = 3 pt, line width = 2.6 pt] plot ({0.5 * sin(\t) - 0.5}, {cos(\t) + 0.3 * (3.14 / 180) * \t});
	
		\draw [<->] (1, {0.3 * 3.14 / 2}) -- node [above] {RI} (2, {0.3 * 3.14 / 2});
	
		\draw (3.5, {-0.3 * 3.14 / 2}) -- (3.5, {0.3 * 3 * 3.14 / 2});

	\end{scope}

\\


	\begin{scope}[xshift = -3 cm]

		\draw (0, {0.3 * 3.14 / 2}) -- (2, {0.3 * 3.14 / 2});
		\draw [draw = white, double = black, double distance between line centers = 3 pt, line width = 2.6 pt] (0.5, {-0.15 * 3.14 / 2}) -- (0.5, {0.3 * 3.14 / 2}) arc (180 : 0 : 0.5) -- (1.5, {-0.15 * 3.14 / 2});
	
		\draw [<->] (2.5, {0.3 * 3.14 / 2}) -- node [above] {RII} (3.5, {0.3 * 3.14 / 2});
	
		\draw (4, 0.75) -- (6, 0.75);
		\draw (4.5, {-0.15 * 3.14 / 2}) arc (180 : 0 : 0.5);

	\end{scope}

\\
	
	
	\draw (-3, 1) -- (-1, -1);
	\draw [draw = white, double = black, double distance between line centers = 3 pt, line width = 2.6 pt]
		plot [variable = \y, samples = 15, domain = -1: 1] ({-0.5 / (1 + exp(-5 * \y)) - 2}, 0.5 * \y - 0.5);
	\draw [draw = white, double = black, double distance between line centers = 3 pt, line width = 2.6 pt]
		plot [variable = \y, samples = 15, domain = -1: 1] ({0.5 / (1 + exp(-5 * \y)) - 2.5}, 0.5 * \y + 0.5);
	\draw [draw = white, double = black, double distance between line centers = 3 pt, line width = 2.6 pt] (-1, 1) -- (-3, -1);
		
	\draw [<->] (-0.5, 0) -- node [midway, above] {RIII} (0.5, 0);
		
	\draw (3, -1) -- (1, 1);
	\draw [draw = white, double = black, double distance between line centers = 3 pt, line width = 2.6 pt]
		plot [variable = \y, samples = 15, domain = -1: 1] ({0.5 / (1 + exp(-5 * \y)) + 2}, 0.5 * \y - 0.5);
	\draw [draw = white, double = black, double distance between line centers = 3 pt, line width = 2.6 pt]
		plot [variable = \y, samples = 15, domain = -1: 1] ({-0.5 / (1 + exp(-5 * \y)) + 2.5}, 0.5 * \y + 0.5);
	\draw [draw = white, double = black, double distance between line centers = 3 pt, line width = 2.6 pt] (1, -1) -- (3, 1);
	
\\
};
\end{tikzpicture}
\caption{\label{reid}The Reidemeister moves RI, RII, and RIII used in knot theory. Not all variations of the rules are shown in this figure.}
\end{figure}
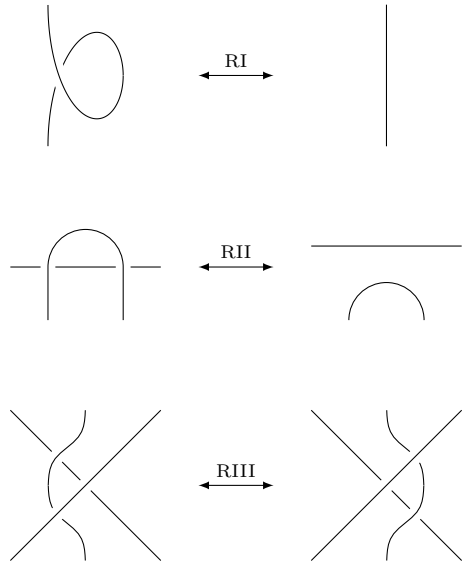

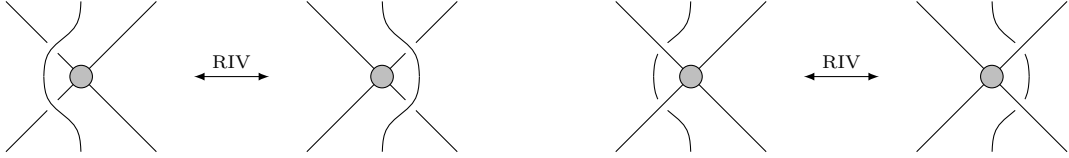
\begin{figure}[hbt]
\centering
\begin{tikzpicture}[> = latex, font = \scriptsize]
\matrix[column sep = 2 cm]{


	\draw (-3, 1) -- (-1, -1) (-1, 1) -- (-3, -1);
	\draw [fill = gray!50] (-2, 0) circle (0.15);

	\draw [draw = white, double = black, double distance between line centers = 3 pt, line width = 2.6 pt]
		plot [variable = \y, samples = 15, domain = -1: 1] ({-0.5 / (1 + exp(-5 * \y)) - 2}, 0.5 * \y - 0.5);
	\draw [draw = white, double = black, double distance between line centers = 3 pt, line width = 2.6 pt]
		plot [variable = \y, samples = 15, domain = -1: 1] ({0.5 / (1 + exp(-5 * \y)) - 2.5}, 0.5 * \y + 0.5);
		
	\draw [<->] (-0.5, 0) -- node [midway, above] {RIV} (0.5, 0);
		
	\draw (3, -1) -- (1, 1) (1, -1) -- (3, 1);
	\draw [fill = gray!50] (2, 0) circle (0.15);

	\draw [draw = white, double = black, double distance between line centers = 3 pt, line width = 2.6 pt]
		plot [variable = \y, samples = 15, domain = -1: 1] ({0.5 / (1 + exp(-5 * \y)) + 2}, 0.5 * \y - 0.5);
	\draw [draw = white, double = black, double distance between line centers = 3 pt, line width = 2.6 pt]
		plot [variable = \y, samples = 15, domain = -1: 1] ({-0.5 / (1 + exp(-5 * \y)) + 2.5}, 0.5 * \y + 0.5);

&


	\draw plot [variable = \y, samples = 15, domain = -1: 1] ({-0.5 / (1 + exp(-5 * \y)) - 2}, 0.5 * \y - 0.5);
	\draw plot [variable = \y, samples = 15, domain = -1: 1] ({0.5 / (1 + exp(-5 * \y)) - 2.5}, 0.5 * \y + 0.5);

	\draw [draw = white, double = black, double distance between line centers = 3 pt, line width = 2.6 pt] (-3, 1) -- (-1, -1) (-1, 1) -- (-3, -1);
	\draw [fill = gray!50] (-2, 0) circle (0.15);
		
	\draw [<->] (-0.5, 0) -- node [midway, above] {RIV} (0.5, 0);

	\draw plot [variable = \y, samples = 15, domain = -1: 1] ({0.5 / (1 + exp(-5 * \y)) + 2}, 0.5 * \y - 0.5);
	\draw plot [variable = \y, samples = 15, domain = -1: 1] ({-0.5 / (1 + exp(-5 * \y)) + 2.5}, 0.5 * \y + 0.5);

	\draw [draw = white, double = black, double distance between line centers = 3 pt, line width = 2.6 pt] (3, 1) -- (1, -1) (1, 1) -- (3, -1);
	\draw [fill = gray!50] (2, 0) circle (0.15);

\\
};
\end{tikzpicture}
\caption{\label{RIV-move}The RIV graph equivalence move.}
\end{figure}

From knot theory, there are a well-known set of equivalence moves, known as the Reidemeister moves, acting on diagrams of knots. These are given in Figure \ref{reid}. Starting from a given diagram of a knot, any other diagram that can be obtained by a finite set of these moves is in the same knot equivalence class. In addition, as shown by Kauffman~\cite{Kau89}, there are additional moves dealing with the presence of vertices in the graph diagram. We shall often refer to these graph equivalence moves as ``Reidemeister moves" as well. Figure \ref{RIV-move} shows the RIV move, where edges may freely pass either over or under a vertex without issue. However, Kaufmann also distinguishes a final move, called the RV move, based on whether the vertices of the graph are considered ``rigid" or ``non-rigid". A rigid vertex can be thought of as a disk with edges attached, and the move is that of flipping the coin over to its opposite side. For the non-rigid vertex, the move is switching the positions of two edges on the vertex. In other words, it allows for the arbitrary permutation of the points of contact for the edges on the vertex. As we will see in Section \ref{vertex}, the stronger condition of diffeomorphism invariance placed on the graphs considered here will allow for a third case, which is part way between these rigid and non-rigid moves.

\subsection{Vertex states}
\label{vertex}

\begin{figure}[hbt]
\centering
\begin{tikzpicture}
\matrix[column sep = 2 cm]{


	\node [cylinder, draw, minimum height = 0.5 cm, minimum width = 0.05 cm, aspect = 0.5,
		left color = gray!70, right color = gray, middle color = gray!50, label = {left : $A$}] at ({-sqrt(8) / 6}, 0) {};

	\node [cylinder, draw, minimum height = 0.5 cm, minimum width = 0.05 cm, aspect = 0.5, rotate = 180,
		left color = gray!70, right color = gray, middle color = gray!50, label = {left : $C$}] at ({sqrt(8) / 6}, 0) {};

	\node [cylinder, draw, minimum height = 0.5 cm, minimum width = 0.05 cm, aspect = 0.5, rotate = -90,
		left color = gray!70, right color = gray, middle color = gray!50] at (0, {-sqrt(8) / 6}) {};

	\draw [ball color = gray!50] (0, 0, 0) circle (0.5);

	\node [cylinder, draw, minimum height = 0.5 cm, minimum width = 0.05 cm, aspect = 0.5, rotate = 90,
		left color = gray!70, right color = gray, middle color = gray!50, label = {right : $B$}] at (0, {sqrt(8) / 6}) {};

	\node [cylinder, draw, minimum height = 0.5 cm, minimum width = 0.05 cm, aspect = 0.5, rotate = -90,
		left color = gray!70, right color = gray, middle color = gray!50, label = {right : $D$}] at (0, {-sqrt(8) / 6}) {};


	\draw (2.5, 0) node [left] {$A$} -- (3.5, 0) node [right] {$C$};
	\draw [fill = white] (3, 0) circle (0.15);
	\draw (3, -0.5) node [below] {$D$} -- (3, 0.5) node [above] {$B$};


	\node at (1.5, -1) {$\oplus$ vertex};

&


	\node [cylinder, draw, minimum height = 0.5 cm, minimum width = 0.05 cm, aspect = 0.5, rotate = -90,
		left color = gray!70, right color = gray, middle color = gray!50, label = {left : $B$}] at (0, {sqrt(8) / 6}) {};

	\node [cylinder, draw, minimum height = 0.5 cm, minimum width = 0.05 cm, aspect = 0.5, rotate = 90,
		left color = gray!70, right color = gray, middle color = gray!50, label = {left : $D$}] at (0, {-sqrt(8) / 6}) {};

	\draw [ball color = gray!50] (0, 0, 0) circle (0.5);

	\node [cylinder, draw, minimum height = 0.5 cm, minimum width = 0.05 cm, aspect = 0.5, rotate = 180,
		left color = gray!70, right color = gray, middle color = gray!50, label = {right : $A$}] at ({-sqrt(8) / 6}, 0) {};

	\node [cylinder, draw, minimum height = 0.5 cm, minimum width = 0.05 cm, aspect = 0.5,
		left color = gray!70, right color = gray, middle color = gray!50, label = {right : $C$}] at ({sqrt(8) / 6}, 0) {};


	\draw (3, -0.5) node [below] {$D$} -- (3, 0.5) node [above] {$B$};
	\draw [fill = white] (3, 0) circle (0.15);
	\draw (2.5, 0) node [left] {$A$} -- (3.5, 0) node [right] {$C$};


	\node at (1.5, -1) {$\ominus$ vertex};

\\
};
\end{tikzpicture}
\caption{\label{vert-states}Physical and schematic representations of the two possible vertex states. For the left-hand picture of each pair, the vertex and its incident edges are shown with finite size, to demonstrate how they connect together.}
\end{figure}
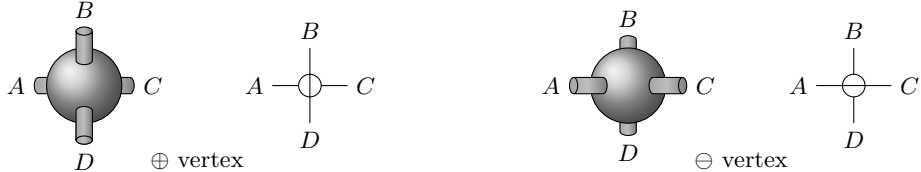

As discussed in Section \ref{intro}, the introduction of diffeomorphism equivalence classes of knotted graphs means that special care must be taken to specify the vertices. In particular, the tangent vectors of the edges incident to a given vertex are chosen so that no three of them are linearly dependent. This prevents the need for a continuous parameter to describe the vertex, and can be done for any valence of four or less~\cite{Gro-Rov96}. Thus, for each such non-degenerate vertex, the edges are incident to the vertex in a tetrahedral pattern. In other words, the edges are placed around the vertex in positions dual to the faces of a tetrahedron. To maintain linear independence of any three edges, no edge can be moved through the plane formed by two of the other edges. When projecting the graph onto a plane to get its graph diagram, the vertices can be rotated as appropriate so that two of the edges are incident above the plane, and two below. To show this in the graph diagram, vertices are labeled with one of two vertex states, $\oplus$ and $\ominus$; this notation was first used by Wan~\cite{Wan07}. Figure \ref{vert-states} shows the physical representation of the vertex for each of the two states, along with the symbol used in the graph diagrams.

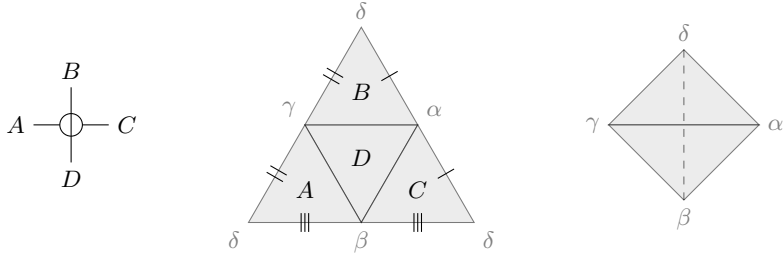
\begin{figure}[hbt]
\centering
\begin{tikzpicture}
\matrix[column sep = 1 cm]{


	\draw (2.5, 0) node [left] {$A$} -- (3.5, 0) node [right] {$C$};
	\draw [fill = white] (3, 0) circle (0.15);
	\draw (3, -0.5) node [below] {$D$} -- (3, 0.5) node [above] {$B$};

&

	\begin{scope}[yshift = -0.433 cm]		

	
		\draw [fill = gray!30, semitransparent] (30 : {sqrt(3) / 2}) node [above right] {$\alpha$} -- (150 : {sqrt(3) / 2}) node [above left] {$\gamma$}
			-- (270 : {sqrt(3) / 2}) node [below] {$\beta$} -- cycle;
		\draw [fill = gray!30, semitransparent] (210 : {sqrt(3)}) node [below left] {$\delta$} -- (150 : {sqrt(3) / 2}) -- (270 : {sqrt(3) / 2}) -- cycle;
		\draw [fill = gray!30, semitransparent] (30 : {sqrt(3) / 2}) -- (270 : {sqrt(3) / 2}) -- (330 : {sqrt(3)}) node [below right] {$\delta$} -- cycle;
		\draw [fill = gray!30, semitransparent] (30 : {sqrt(3) / 2}) -- (150 : {sqrt(3) / 2}) -- (90 : {sqrt(3)}) node [above] {$\delta$} -- cycle;
	
	
		\node at (0, 0) {$D$};
		\node at (90 : {sqrt(3) / 2}) {$B$};
		\node at (210 : {sqrt(3) / 2}) {$A$};
		\node at (330 : {sqrt(3) / 2}) {$C$};
	

	
		\draw ({-sqrt(3) / 4 * cos(30) - 0.125 * cos(30) + 0.05 * cos(60)}, {sqrt(3) / 2 + sqrt(3) / 4 * sin(30) + 0.125 * sin(30) + 0.05 * sin(60)}) -- 
			({-sqrt(3) / 4 * cos(30) + 0.125 * cos(30) + 0.05 * cos(60)}, {sqrt(3) / 2 + sqrt(3) / 4 * sin(30) - 0.125 * sin(30) + 0.05 * sin(60)});
		\draw ({-sqrt(3) / 4 * cos(30) - 0.125 * cos(30) - 0.05 * cos(60)}, {sqrt(3) / 2 + sqrt(3) / 4 * sin(30) + 0.125 * sin(30) - 0.05 * sin(60)}) -- 
			({-sqrt(3) / 4 * cos(30) + 0.125 * cos(30) - 0.05 * cos(60)}, {sqrt(3) / 2 + sqrt(3) / 4 * sin(30) - 0.125 * sin(30) - 0.05 * sin(60)});

		\draw ({-sqrt(3) / 2 * cos(330) - sqrt(3) / 4 * cos(30) - 0.125 * cos(30) + 0.05 * cos(60)}, {sqrt(3) / 2 * sin(330) + sqrt(3) / 4 * sin(30) + 0.125 * sin(30) + 0.05 * sin(60)}) --
			({-sqrt(3) / 2 * cos(330) - sqrt(3) / 4 * cos(30) + 0.125 * cos(30) + 0.05 * cos(60)}, {sqrt(3) / 2 * sin(330) + sqrt(3) / 4 * sin(30) - 0.125 * sin(30) + 0.05 * sin(60)});
		\draw ({-sqrt(3) / 2 * cos(330) - sqrt(3) / 4 * cos(30) - 0.125 * cos(30) - 0.05 * cos(60)}, {sqrt(3) / 2 * sin(330) + sqrt(3) / 4 * sin(30) + 0.125 * sin(30) - 0.05 * sin(60)}) --
			({-sqrt(3) / 2 * cos(330) - sqrt(3) / 4 * cos(30) + 0.125 * cos(30) - 0.05 * cos(60)}, {sqrt(3) / 2 * sin(330) + sqrt(3) / 4 * sin(30) - 0.125 * sin(30) - 0.05 * sin(60)});

	
		\draw ({sqrt(3) / 4 * cos(30) + 0.125 * cos(30)}, {sqrt(3) / 2 + sqrt(3) / 4 * sin(30) + 0.125 * sin(30)}) --
			({sqrt(3) / 4 * cos(30) - 0.125 * cos(30)}, {sqrt(3) / 2 + sqrt(3) / 4 * sin(30) - 0.125 * sin(30)});

		\draw ({sqrt(3) / 2 * cos(330) + sqrt(3) / 4 * cos(30) + 0.125 * cos(30)}, {sqrt(3) / 2 * sin(330) + sqrt(3) / 4 * sin(30) + 0.125 * sin(30)}) --
			({sqrt(3) / 2 * cos(330) + sqrt(3) / 4 * cos(30) - 0.125 * cos(30)}, {sqrt(3) / 2 * sin(330) + sqrt(3) / 4 * sin(30) - 0.125 * sin(30)});

	
		\draw (-0.7, {-sqrt(3) / 2 + 0.125}) -- (-0.7, {-sqrt(3) / 2 - 0.125});
		\draw (-0.75, {-sqrt(3) / 2 + 0.125}) -- (-0.75, {-sqrt(3) / 2 - 0.125});
		\draw (-0.8, {-sqrt(3) / 2 + 0.125}) -- (-0.8, {-sqrt(3) / 2 - 0.125});
	
		\draw (0.7, {-sqrt(3) / 2 + 0.125}) -- (0.7, {-sqrt(3) / 2 - 0.125});
		\draw (0.75, {-sqrt(3) / 2 + 0.125}) -- (0.75, {-sqrt(3) / 2 - 0.125});
		\draw (0.8, {-sqrt(3) / 2 + 0.125}) -- (0.8, {-sqrt(3) / 2 - 0.125});

	\end{scope}

&

	\draw [dashed] (0, 1) -- (0, -1);
	\draw [fill = gray!30, semitransparent] (-1, 0) node [left] {$\gamma$} -- (1, 0) node [right] {$\alpha$} -- (0, 1) node [above] {$\delta$} -- cycle;
	\draw [fill = gray!30, semitransparent] (-1, 0) -- (1, 0) -- (0, -1) node [below] {$\beta$} -- cycle;

\\
};
\end{tikzpicture}
\caption{\label{plus-dual-state}The $\oplus$ vertex state, and the corresponding tetrahedron in the dual simplicial complex.}
\end{figure}

\begin{figure}[hbt]
\centering
\begin{tikzpicture}
\matrix[column sep = 1 cm]{


	\draw (3, -0.5) node [below] {$D$} -- (3, 0.5) node [above] {$B$};
	\draw [fill = white] (3, 0) circle (0.15);
	\draw (2.5, 0) node [left] {$A$} -- (3.5, 0) node [right] {$C$};

&

	\begin{scope}[rotate = 90]		

	
		\draw [fill = gray!30, semitransparent] (30 : {sqrt(3) / 2}) node [above left] {$\delta$} -- (150 : {sqrt(3) / 2}) node [below left] {$\beta$}
			-- (270 : {sqrt(3) / 2}) node [right] {$\alpha$} -- cycle;
		\draw [fill = gray!30, semitransparent] (210 : {sqrt(3)}) node [below left] {$\gamma$} -- (150 : {sqrt(3) / 2}) -- (270 : {sqrt(3) / 2}) -- cycle;
		\draw [fill = gray!30, semitransparent] (30 : {sqrt(3) / 2}) -- (270 : {sqrt(3) / 2}) -- (330 : {sqrt(3)}) node [above right] {$\gamma$} -- cycle;
		\draw [fill = gray!30, semitransparent] (30 : {sqrt(3) / 2}) -- (150 : {sqrt(3) / 2}) -- (90 : {sqrt(3)}) node [above] {$\gamma$} -- cycle;
	
	
		\node at (0, 0) {$C$};
		\node at (90 : {sqrt(3) / 2}) {$A$};
		\node at (210 : {sqrt(3) / 2}) {$D$};
		\node at (330 : {sqrt(3) / 2}) {$B$};
	

	
		\draw ({-sqrt(3) / 4 * cos(30) - 0.125 * cos(30) + 0.05 * cos(60)}, {sqrt(3) / 2 + sqrt(3) / 4 * sin(30) + 0.125 * sin(30) + 0.05 * sin(60)}) -- 
			({-sqrt(3) / 4 * cos(30) + 0.125 * cos(30) + 0.05 * cos(60)}, {sqrt(3) / 2 + sqrt(3) / 4 * sin(30) - 0.125 * sin(30) + 0.05 * sin(60)});
		\draw ({-sqrt(3) / 4 * cos(30) - 0.125 * cos(30) - 0.05 * cos(60)}, {sqrt(3) / 2 + sqrt(3) / 4 * sin(30) + 0.125 * sin(30) - 0.05 * sin(60)}) -- 
			({-sqrt(3) / 4 * cos(30) + 0.125 * cos(30) - 0.05 * cos(60)}, {sqrt(3) / 2 + sqrt(3) / 4 * sin(30) - 0.125 * sin(30) - 0.05 * sin(60)});

		\draw ({-sqrt(3) / 2 * cos(330) - sqrt(3) / 4 * cos(30) - 0.125 * cos(30) + 0.05 * cos(60)}, {sqrt(3) / 2 * sin(330) + sqrt(3) / 4 * sin(30) + 0.125 * sin(30) + 0.05 * sin(60)}) --
			({-sqrt(3) / 2 * cos(330) - sqrt(3) / 4 * cos(30) + 0.125 * cos(30) + 0.05 * cos(60)}, {sqrt(3) / 2 * sin(330) + sqrt(3) / 4 * sin(30) - 0.125 * sin(30) + 0.05 * sin(60)});
		\draw ({-sqrt(3) / 2 * cos(330) - sqrt(3) / 4 * cos(30) - 0.125 * cos(30) - 0.05 * cos(60)}, {sqrt(3) / 2 * sin(330) + sqrt(3) / 4 * sin(30) + 0.125 * sin(30) - 0.05 * sin(60)}) --
			({-sqrt(3) / 2 * cos(330) - sqrt(3) / 4 * cos(30) + 0.125 * cos(30) - 0.05 * cos(60)}, {sqrt(3) / 2 * sin(330) + sqrt(3) / 4 * sin(30) - 0.125 * sin(30) - 0.05 * sin(60)});

	
		\draw ({sqrt(3) / 4 * cos(30) + 0.125 * cos(30)}, {sqrt(3) / 2 + sqrt(3) / 4 * sin(30) + 0.125 * sin(30)}) --
			({sqrt(3) / 4 * cos(30) - 0.125 * cos(30)}, {sqrt(3) / 2 + sqrt(3) / 4 * sin(30) - 0.125 * sin(30)});

		\draw ({sqrt(3) / 2 * cos(330) + sqrt(3) / 4 * cos(30) + 0.125 * cos(30)}, {sqrt(3) / 2 * sin(330) + sqrt(3) / 4 * sin(30) + 0.125 * sin(30)}) --
			({sqrt(3) / 2 * cos(330) + sqrt(3) / 4 * cos(30) - 0.125 * cos(30)}, {sqrt(3) / 2 * sin(330) + sqrt(3) / 4 * sin(30) - 0.125 * sin(30)});

	
		\draw (-0.7, {-sqrt(3) / 2 + 0.125}) -- (-0.7, {-sqrt(3) / 2 - 0.125});
		\draw (-0.75, {-sqrt(3) / 2 + 0.125}) -- (-0.75, {-sqrt(3) / 2 - 0.125});
		\draw (-0.8, {-sqrt(3) / 2 + 0.125}) -- (-0.8, {-sqrt(3) / 2 - 0.125});
	
		\draw (0.7, {-sqrt(3) / 2 + 0.125}) -- (0.7, {-sqrt(3) / 2 - 0.125});
		\draw (0.75, {-sqrt(3) / 2 + 0.125}) -- (0.75, {-sqrt(3) / 2 - 0.125});
		\draw (0.8, {-sqrt(3) / 2 + 0.125}) -- (0.8, {-sqrt(3) / 2 - 0.125});

	\end{scope}

&

	\draw [dashed] (-1, 0) -- (1, 0);
	\draw [fill = gray!30, semitransparent] (-1, 0) node [left] {$\gamma$} -- (0, 1) node [above] {$\delta$} -- (0, -1) node [below] {$\beta$} -- cycle;
	\draw [fill = gray!30, semitransparent] (1, 0) node [right] {$\alpha$} -- (0, 1) -- (0, -1) -- cycle;

\\
};
\end{tikzpicture}
\caption{\label{minus-dual-state}The $\ominus$ vertex state, and the corresponding tetrahedron in the dual simplicial complex.}
\end{figure}
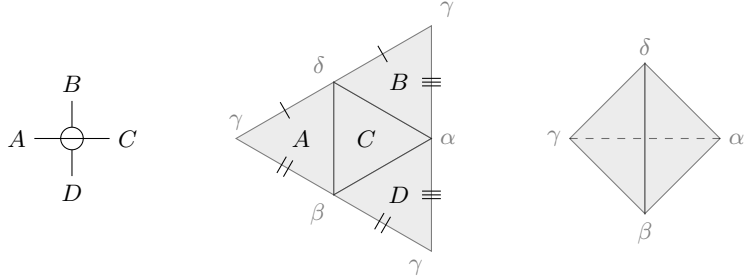

At this point, we will address the difference between the two vertex states $\oplus$ and $\ominus$. To do this, we will consider a single vertex, along with its vertex state and the edges incident to the vertex. For every such vertex, there is a dual tetrahedron, where the edges of the graph correspond to the faces of the tetrahedron. This correspondence is shown in Figure \ref{plus-dual-state} for the $\oplus$ vertex state, and Figure \ref{minus-dual-state} for the $\ominus$ state. We focus first on the $\oplus$ state and Figure \ref{plus-dual-state}. The original vertex is shown on the left-hand side of the figure, where each edge is given a label $\{A, B, C, D\}$; the dual tetrahedron is shown on the right-hand side of the figure, where each corner of the tetrahedron is given a label $\{\alpha, \beta, \gamma, \delta \}$. Thus, the faces of the tetrahedron can be given labels from both the three corners of the face, and from the edge in the dual graph. The correspondence between the edge labels of the vertex, and the face labels of the tetrahedron, are shown in the middle of Figure \ref{plus-dual-state}, where the faces of the tetrahedron have been projected onto the plane. Thus, for example, edge $A$ of the graph is matched with face $\beta \gamma \delta$ of the tetrahedron. Note that the projection in the middle of Figure \ref{plus-dual-state} is done so that the faces shown (i.e. facing the reader on the plane of projection) are the exterior faces. In this projection, the faces around the middle face $D$ are $B, A, C$, going in counter-clockwise order.

Now we use the same process on a vertex in the $\ominus$ state, as shown in Figure \ref{minus-dual-state}. Here, it is assumed that the rest of the graph has remained the same, so in particular, the four edges $\{A, B, C, D\}$ have the same connections as before. We will see how changing the vertex state, with no other changes in the graph diagram, leads to a different situation. For the left-hand diagram showing the vertex, this means the edges $A, C$ now point above the projection plane of the vertex, while edges $B, D$ point below it. This leads to a corresponding change in the dual tetrahedron, shown to the right of the figure. There, faces $\alpha \beta \delta, \beta \gamma \delta$ matching edges $A, C$ also are above the plane of projection, and faces $\alpha \beta \gamma, \alpha \gamma \delta$ (dual to edges $D, B$, respectively), are below the plane. Now the face $\alpha \beta \delta$ is chosen as the central face for the middle diagram of Figure \ref{minus-dual-state}, so that the faces represented there are again the exterior faces of the dual tetrahedron. However, note that there has been a relative change in the faces. In particular, again going around face $D$ in counter-clockwise order, the other faces are now arranged as $B, C, A$.  This corresponds to a tetrahedron with its orientation reversed from that shown in Figure \ref{plus-dual-state}. Thus, changing the vertex state of a particular vertex is equivalent to replacing a dual tetrahedron in a triangulation with one having the opposite orientations on its faces.

\subsection{Vertex rotations}
\label{vert-rot}

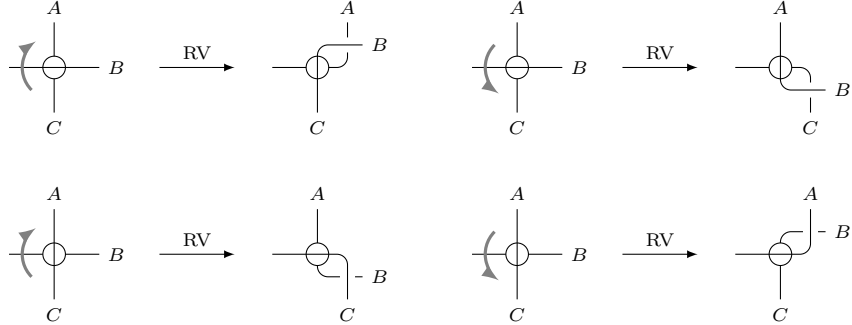
\begin{figure}[hbt]
\centering
\begin{tikzpicture}[> = latex, font = \scriptsize]
\matrix[column sep = 1 cm, row sep = 0.5 cm]{


	\draw (0, -0.6) node [below] {$C$} -- (0, 0.6) node [above] {$A$};
	\draw [fill = white] (0, 0) circle (0.15);
	\draw (-0.6, 0) -- (0.6, 0) node [right] {$B$};

	\draw [->, very thick, gray] (-0.3, -0.3) arc (225 : 125 : {0.3 * sqrt(2)});

	\draw [->] (1.4, 0) -- node [above] {RV} (2.4, 0);


	\draw [rounded corners] (2.9, 0) -- (3.9, 0) -- (3.9, 0.2) (3.9, 0.4) -- (3.9, 0.6) node [above] {$A$};
	\draw [fill = white] (3.5, 0) circle (0.15);
	\draw [rounded corners] (3.5, -0.6) node [below] {$C$} -- (3.5, 0.3) -- (4.1, 0.3) node [right] {$B$};

&


	\draw (0, -0.6) node [below] {$C$} -- (0, 0.6) node [above] {$A$};
	\draw [fill = white] (0, 0) circle (0.15);
	\draw (-0.6, 0) -- (0.6, 0) node [right] {$B$};

	\draw [->, very thick, gray] (-0.3, 0.3) arc (135 : 235 : {0.3 * sqrt(2)});

	\draw [->] (1.4, 0) -- node [above] {RV} (2.4, 0);


	\draw [rounded corners] (2.9, 0) -- (3.9, 0) -- (3.9, -0.2) (3.9, -0.4) -- (3.9, -0.6) node [below] {$C$};
	\draw [fill = white] (3.5, 0) circle (0.15);
	\draw [rounded corners] (3.5, 0.6) node [above] {$A$} -- (3.5, -0.3) -- (4.1, -0.3) node [right] {$B$};

\\


	\draw (-0.6, 0) -- (0.6, 0) node [right] {$B$};
	\draw [fill = white] (0, 0) circle (0.15);
	\draw (0, -0.6) node [below] {$C$} -- (0, 0.6) node [above] {$A$};

	\draw [->, very thick, gray] (-0.3, -0.3) arc (225 : 125 : {0.3 * sqrt(2)});

	\draw [->] (1.4, 0) -- node [above] {RV} (2.4, 0);


	\draw [rounded corners] (3.5, 0.6) node [above] {$A$} -- (3.5, -0.3) -- (3.8, -0.3) (4, -0.3) -- (4.1, -0.3) node [right] {$B$};
	\draw [fill = white] (3.5, 0) circle (0.15);
	\draw [rounded corners] (2.9, 0) -- (3.9, 0) -- (3.9, -0.6) node [below] {$C$};

&


	\draw (-0.6, 0) -- (0.6, 0) node [right] {$B$};
	\draw [fill = white] (0, 0) circle (0.15);
	\draw (0, -0.6) node [below] {$C$} -- (0, 0.6) node [above] {$A$};

	\draw [->, very thick, gray] (-0.3, 0.3) arc (135 : 235 : {0.3 * sqrt(2)});

	\draw [->] (1.4, 0) -- node [above] {RV} (2.4, 0);


	\draw [rounded corners] (3.5, -0.6) node [below] {$C$} -- (3.5, 0.3) -- (3.8, 0.3) (4, 0.3) -- (4.1, 0.3) node [right] {$B$};
	\draw [fill = white] (3.5, 0) circle (0.15);
	\draw [rounded corners] (2.9, 0) -- (3.9, 0) -- (3.9, 0.6) node [above] {$A$};

\\
};

\end{tikzpicture}
\caption{\label{RV-move}The RV move for rotating a vertex around one of its incident edges, based on its current vertex state. The vertex is rotated $\pi / 3$ around the left edge in the direction shown by the arrow in each case. This rotation will change the vertex to the opposite vertex state, and create a crossing between two of the edges incident to the vertex. The top row shows how rotations change a $\ominus$ state into a $\oplus$ state, while the bottom row shows the reverse.}
\end{figure}

Now that the vertex states $\oplus$ and $\ominus$ have been defined, we turn to the possible rotations of the vertices. Recall that in Section \ref{graph-def}, vertex rotations for rigid and non-rigid graphs, as given by Kauffman~\cite{Kau89}, were described. Here, the RV move on vertices will be defined differently. The key issue is that the tetrahedral structure of the edges incident to the vertex must be preserved under such rotations. Because of this, it is not possible to freely permute any two edges, as would be the case for a non-rigid vertex. However, half of these permutations will be possible, depending on the vertex state of the vertex being rotated. This allows more freedom for such moves that the rigid vertex case of Kauffman, but not using any pair of edges adjacent in the graph diagram, as with the non-rigid case. Now, consider a rotation of the vertex around an axis along the tangent vector of one of incident edges. The minimal rotation of $\pi/3$ around the axis takes a vertex from one vertex state to another~\cite{Wan07}. This will be our definition of the RV move. Two of these moves in the same direction takes the vertex back to its original state, although the edges incident to the vertex will have changes in the number of crossings between them. A sample of these rotations around one of the four edges is shown in Figure \ref{RV-move}; all of the other possibilities are given by appropriate rotations of these diagrams.

This now completes the list of graph equivalence moves. To summarize, the moves RI, RII, and RIII are the same as those in knot theory. The RIV move (edges passing over or under a vertex) and the RV move (vertex rotations) deal with vertices. We can now list a few simple invariants of knotted graphs. Obviously, none of these moves add or delete vertices in the graph, so the vertex number stays the same.

\begin{proposition}
	The number of vertices of a knotted 4-regular pseudograph is an invariant under the graph equivalence moves.
\end{proposition}
In addition, neither the RIV nor the RV move changes the edges incident to a given vertex. Thus, the connectivity of the graph is invariant under these moves. We can state this in several ways, but here, it is given in terms of the adjacency matrix $A_{ij}$, where $i, j$ are (arbitrary) labels for vertices. $A_{ij} = 0$ if there is no connection between vertices $i$ and $j$, while $A_{ij}$ gives the number of distinct connections when $i \ne j$. Finally $A_{ii} = 2$ for a self-loop in a graph with more than one vertex, and $A_{ii} = 0$ otherwise.

\begin{proposition}
	The adjacency matrix of a knotted 4-regular pseudograph is an invariant under the graph equivalence moves.
\end{proposition}
There is one implication of this invariant that will be important later in this work. Recall that an Eulerian circuit is a path through a graph that begins and ends at the same vertex, and passes over each edge exactly once. Such a circuit exists for a graph if and only if all of the vertices have an even valence, which is satisfied by the graphs under consideration here. The Eulerian circuit is thus a sequence of (oriented) edges in the graph. After any graph equivalence move, this sequence is still valid, although the manner in which the circuit passes through any particular vertex may change. Because the adjacency matrix of the graph is invariant under such moves, the number of Eulerian circuits depends only on the abstract graph, and not its embedding. This gives our next invariant.

\begin{corollary}
	The number of Eulerian cycles of a knotted 4-regular pseudograph is an invariant under the graph equivalence moves.
\end{corollary}

\section{Graph nomenclature}
\label{notation}

\begin{figure}[hbt]
\centering
\begin{tikzpicture}[> = latex, font = \footnotesize]
	
	
	\def\d{1}
	
	
	\node [label = 215 : 1, label = 45 : 4] at (-\d, 2 * \d) {};
	\node [label = 215 : 2, label = 45 : 7] (node-27) at (0, 2 * \d) {};
	\node [label = 215 : 3, label = 45 : 6] (node-36) at (\d, 2 * \d) {};
	\node [label = 215 : 5, label = 45 : 8] at (-\d, \d) {};
	
	\draw [rounded corners] (node-27.east) -- (-2 * \d, 2 * \d) -- (-2 * \d, 1.75 * \d);
	
	\begin{scope}[decoration = {markings, mark = at position 0.5 with {\arrow{latex}}}]
	
		\draw [postaction = {decorate}] (node-27.east) -- (node-36.west);
		\draw [postaction = {decorate}, rounded corners] (node-36.north) -- (\d, 2.75 * \d) -- (0, 2.75 * \d) -- (node-27.north);
		
		\draw [postaction = {decorate}] (-2 * \d, 1.25 * \d) -- (-2 * \d, 1.75 * \d);
		\draw [postaction = {decorate}] (-0.5 * \d, 0.5 * \d) -- (0.5 * \d, 0.5 * \d);
		\draw [postaction = {decorate}] (\d, 0.75 * \d) -- (node-36.north);
		\draw [postaction = {decorate}] (1.5 * \d, 3 * \d) -- (-0.5 * \d, 3 * \d);
	
	\end{scope}
	
	\begin{scope}[rounded corners]
	
		\draw [draw = white, double = black, double distance between line centers = 3 pt, line width = 2.6 pt] (-0.5 * \d, 3 * \d) -- (-\d, 3 * \d) -- (-\d, 0.5 * \d) -- (-0.5 * \d, 0.5 * \d);
	
		\draw [draw = white, double = black, double distance between line centers = 3 pt, line width = 2.6 pt] (-2 * \d, 1.25 * \d) -- (-2 * \d, \d) -- (0, \d) -- (node-27.south);
	
		\draw (node-36.east) -- (2 * \d, 2 * \d) -- (2 * \d, 3 * \d) -- (1.5 * \d, 3 * \d);
		\draw (0.5 * \d, 0.5 * \d) -- (\d, 0.5 * \d) -- (\d, 0.75 * \d);
	
	\end{scope}

\end{tikzpicture}
\caption{\label{knot}A visual representation of the knot given by the DT sequence 4 6 8 2.}
\end{figure}
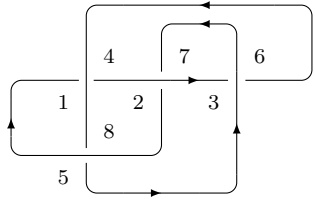

The nomenclature used for the graphs presented here is based on the work of Dowker and Thistlethwaite~\cite{Dow-Thi83}, for enumerating knots. It is useful to summarize their scheme here, to help understand the corresponding version for graphs. Any diagram of a knot is a 4-regular graph, with the nodes composed only of crossings. Because every knot is an embedding of the circle $S^1$, there is an oriented circuit through the knot. In terms of the 4-regular graph, this is an Eulerian circuit, which at every node passes from one edge to the edge directly across from it. Now, suppose one starts at any given point on the graph that is not a node, and moves in the direction of the circuit. As one encounters a crossing, number it with a label, starting with the number 1. Since each crossing is encounted twice, it will receive two numeric labels (one odd and one even). This continues until the starting point is reached, which completes the circuit. A {\it Dowker-Thistletwaite (DT) sequence} is a list of the even labels, given in terms of their matching odd labels. An example is shown in Figure \ref{knot}, for the sequence 4 6 8 2. This example is an alternating knot, where under- and overcrossings alternate. When this is not true of the diagram, a negative sign is added to the number if the crossing is labeled by the even number as an undercrossing is reached. Note that the crossing labels $\ell$ can be shifted by transformations of the form $\ell \to \ell + b$ or $\ell \to b - \ell$, for $1 \le b \le (2N - 1)$ for an $N$-crossing knot, without affecting any of the conditions given above. The sequence that comes in lowest numerical order is chosen to be the representative of this equivalence class of sequences. These transformations only change the crossing labels for a specific diagram of a link, and do not count the effect of the Reidemeister moves changing one diagram to another. Thus, for any graph diagram with a DT sequence of lowest numerical order, there may be an equivalent diagram with fewer crossings, with its own DT sequence.

A similiar process is used to find a DT sequence to describe a graph diagram. However, now nodes can be either crossings or vertices. Unlike the crossings of knots, there are multiple ways an Eulerian circuit can pass through a vertex. For a particular graph diagram and a chosen Eulerian circuit, if the circuit passes through the vertex from the entering edge to the edge directly across from it, the vertex is said to be {\it proper}; otherwise, it is {\it improper}. If all vertices are proper, then the graph diagram is also said to be proper. For a given graph diagram, most (if not all) Eulerian circuits through the diagram will have improper vertices. However, to define a DT sequence for a given knotted 4-valent graph $G$, it is sufficient to find a single graph diagram of $G$, and a particular Eulerian circuit for that diagram, where the diagram is proper. It is possible to do this using simple transformations of the circuit, and the RV vertex rotation move. We now show how this is done.

\begin{figure}[hbt]
\centering
\begin{tikzpicture}[> = latex,
	decoration = {markings, mark = at position 0.75 with {\arrow{latex}}}]
	
	\matrix[column sep = 0.5 cm]{

	\draw (-0.75, 0) node [left] {$A$} -- (0.75, 0) node [right] {$C$};
	\draw (0, -0.75) node [below] {$D$} -- (0, 0.75) node [above] {$B$};

	\draw [fill = gray!50] (0, 0) circle (0.15) node [above right] {$V$};

	\begin{scope}[->, dotted, rounded corners]

		\draw (0.75, -0.2) -- (0.2, -0.2) -- (0.2, -0.75);
		\draw (-0.75, 0.2) -- (-0.2, 0.2) -- (-0.2, 0.75);

	\end{scope}
	
	&
	
	\draw [->] (-0.5, 0) -- node [midway, above, font = \scriptsize] {Rev. $BVCB$} (0.5, 0);
	
	&

	\draw [postaction = {decorate}] (-0.75, 0) node [left] {$A$} -- (-0.15, 0);
	\draw [postaction = {decorate}] (0, 0.75) node [above] {$B$} -- (0, 0.15);
	\draw [postaction = {decorate}] (0.15, 0) -- (0.75, 0) node [right] {$C$};
	\draw [postaction = {decorate}] (0, -0.15) -- (0, -0.75) node [below] {$D$};

	\draw [fill = gray!50] (0, 0) circle (0.15) node [above right] {$V$};
	
	\\
	};

\end{tikzpicture}
\caption{\label{bad-vert-1}A vertex $V$ where the Eulerian cycle passes through the vertex as shown on the left can be placed in proper form by reversing all the orientations along the loop $BVCB$. The vertex state is irrelevant, so it is not shown.}
\end{figure}
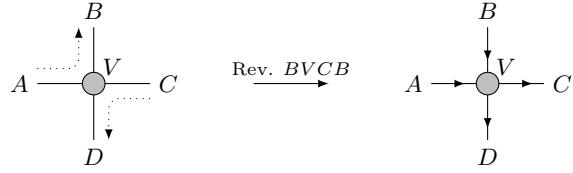

Up to reflection and rotation symmetry, there are two possibilities for improper vertices; for each of these vertices, a method is demonstrated for putting them into proper form. One possibility for such a vertex is shown in Figure \ref{bad-vert-1}. Since the following argument does not depend on the vertex state, it is not shown. In this figure, the Eulerian cycle passes through the vertex $V$ in the order $A \to V \to B$ and $C \to V \to D$. In other words, there is a series of edges along the Eulerian cycle through the rest of the graph between the pair points $A$ and $D$, as well as the pair $B$ and $C$. Thus, there is a sequence of edges in the graph that passes forming a loop $BVCB$, which is not the entire graph $G$. If all the edge orientations on this loop are reversed, this puts the vertex $V$ in proper form. Since the choice of loop is along the original Eulerian cycle, it does not affect the other vertices -- those already in proper form will stay that way, while those still in improper form will remain so, just with two of their incident edges reversed in orientation. Note that reversing some of the edge orientations is not the only graph operation that could have been used; for example, using a RV move on the edges $B$ and $C$, or similarly edges $A$ and $D$, to create a crossing would have put the vertex $V$ in proper form.

\begin{figure}[hbt]
\centering
\begin{tikzpicture}[> = latex,
	decoration = {markings, mark = at position 0.75 with {\arrow{latex}}}]
	
	\matrix[column sep = 0.5 cm]{

	\draw (0, -0.75) node [below] {$D$} -- (0, 0.75) node [above] {$B$};
	\draw [fill = white] (0, 0) circle (0.15) node [above right] {$V$};
	\draw (-0.75, 0) node [left] {$A$} -- (0.75, 0) node [right] {$C$};

	\begin{scope}[->, dotted, rounded corners]

		\draw (0.2, -0.75) -- (0.2, -0.2) -- (0.75, -0.2);
		\draw (-0.75, 0.2) -- (-0.2, 0.2) -- (-0.2, 0.75);

	\end{scope}
	
	&
	
	\draw [->] (-0.5, 0) -- node [midway, above, font = \scriptsize] {RV move} (0.5, 0);
	
	&

	\draw [postaction = {decorate}] (-0.75, 0) node [left] {$A$} -- (-0.15, 0);
	\draw [postaction = {decorate}] (0, 0.15) -- (0, 0.6);
	\draw [postaction = {decorate}] (0.15, 0) -- (0.6, 0);
	\draw [postaction = {decorate}] (0, -0.75) node [below] {$D$} -- (0, -0.15);

	\draw [fill = white] (0, 0) circle (0.15) node [above right] {$V$};
	\draw (0, -0.15) -- (0, 0.15);

	\draw [rounded corners] (0.6, 0) -- (0.75, 0) -- (0.75, 1) node [above] {$B$};
	\draw [rounded corners, draw = white, double = black, double distance between line centers = 3 pt, line width = 2.6 pt] (0, 0.6) -- (0, 0.75) -- (1, 0.75) node [right] {$C$};
	
	\\
	};

\end{tikzpicture}
\caption{\label{bad-vert-2}A vertex $V$ where the Eulerian cycle passes through the vertex as shown on the left can be placed in proper form by using an RV move on the edges $B, C$. Shown here is the case with a $\ominus$ vertex state. A similar RV move is possible for an $\oplus$ vertex as well, resulting in the opposite crossing type.}
\end{figure}
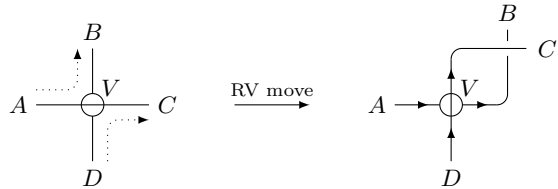

Another possibility for an improper vertex is shown in Figure \ref{bad-vert-2}. Here, the Eulerian cycle passes through the vertex in the order $A \to V \to B$ and $D \to V \to C$. Thus, the remaining cycle joins the pair of points $A$ and $C$, and the pair $B$ and $D$. Unlike the previous case, there are no possible reversals that would put the vertex in proper form. Flipping the orientation of all edges on a closed loop through $V$, but not including the entire graph, does not lead to the vertex becoming proper. In fact, it just leads to a rotated version of the situation on the left of Figure \ref{bad-vert-2}. For example, reversing the loop $AVCA$ results in the Eulerian circuit now following the connections $D \to V \to A$ and $C \to V \to B$. Thus, a RV move is necessary, on two edges whose orientation is either both away from the vertex, or both towards it. In Figure \ref{bad-vert-2}, with a $\ominus$ vertex state, the move is used by rotating on edge $D$, leading to a crossing with edges $B$ and $C$. This is not the only possibility -- the crossing could have been placed on the edges $A$ and $D$. For a $\oplus$, a similar process is needed, although the resulting crossing would have to be different. The Eulerian cycle passes through the vertex as before, with $A \to V \to B$ and $D \to V \to C$, but now the cycle goes through the vertex from one edge to another on the opposite side of the vertex.

Since neither of the two operations in Figures \ref{bad-vert-1} and \ref{bad-vert-2} will take other, already proper vertices and make them improper, continuing this process for all improper vertices will make the graph proper in a finite number of moves. In addition, having a self-loop on the improper vertex does not change this conclusion. For the case in Figure \ref{bad-vert-1}, a self-loop on edges $B, C$ just reverses the orientation on that self-loop. If a self-loop is at the improper vertex in Figure \ref{bad-vert-2}, it cannot have both $B$ and $C$ (or else $A$ and $D$), so the RV move cannot subsequently be undone by a RI move. These considerations prove the following result.
\begin{proposition}
\label{DT-seq-prop}
	For a given choice of Eulerian circuit through the graph diagram of a knotted 4-valent pseudograph, the graph diagram can be made proper by using the orientation reversals and RV moves described above.
\end{proposition}

\begin{figure}[hbt]
\centering
\begin{tikzpicture}[> = latex]
\matrix[column sep = 1 cm]{


	\draw [->] (0, -0.8) node [below] {$D$} -- (0, 0.8) node [above] {$B$};
	\draw [fill = white] (0, 0) circle (0.15);
	\draw [->] (-0.8, 0) node [left] {$A$} -- (0.8, 0) node [right] {$C$};

&


	\draw [thick, ->] (0, -1.25) node [below] {$D$} -- (0, 1.25) node [above] {$B$};


	\node [cylinder, draw, minimum height = 0.5 cm, minimum width = 0.05 cm, aspect = 0.5, rotate = -90,
		left color = gray!70, right color = gray, middle color = gray!50] at (0, {sqrt(8) / 6}) {};

	\node [cylinder, draw, minimum height = 0.5 cm, minimum width = 0.05 cm, aspect = 0.5, rotate = 90,
		left color = gray!70, right color = gray, middle color = gray!50] at (0, {-sqrt(8) / 6}) {};

	\draw [ball color = gray!50] (0, 0) circle (0.5);

	\node [cylinder, draw, minimum height = 0.5 cm, minimum width = 0.05 cm, aspect = 0.5, rotate = 180,
		left color = gray!70, right color = gray, middle color = gray!50] at ({-sqrt(8) / 6}, 0) {};

	\node [cylinder, draw, minimum height = 0.5 cm, minimum width = 0.05 cm, aspect = 0.5,
		left color = gray!70, right color = gray, middle color = gray!50] at ({sqrt(8) / 6}, 0) {};


	\draw [thick, ->] (-1.25, 0) node [left] {$A$} -- (1.25, 0) node [right] {$C$};

\\
};
\end{tikzpicture}
\caption{\label{vert-cross}Eulerian cycle passing along outside of spherical vertex for a $\ominus$ state, projected into the plane.}
\end{figure}
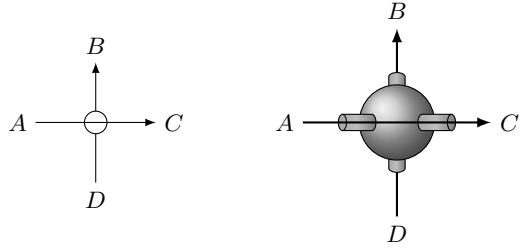

From this result, we can assume from this point forward that the graph diagram can be made proper, and define a DT sequence for graphs similar to how it is done for knots. As with knots, the DT sequence is a series of numbers that indicate the order of nodes as one encounters traveling along an Eulerian cycle through the graph diagram, as well as the type of crossing or vertex as appropriate. In order to distinguish the sequences used for graphs from those for knots, we start the labels from 0 instead of 1, and report the sequence using the odd numbers versus the evens. There is one final point to consider, namely how to designate the vertex state of the graph, in terms of its node labels. We can think of the vertex itself as a small sphere. Because every vertex is now proper, the Eulerian circuit passes from one edge, along the surface of the spherical vertex, and then forward along the opposite edge. An example of this is shown in Figure \ref{vert-cross}. Thus, we can think of one sequence of edges as ``passing" over the other at the vertex, in this particular graph diagram. Using this idea, we now define DT sequences as follows.

\begin{figure}[hbt]
\centering
\begin{tikzpicture}[> = latex, font = \footnotesize]
	
	
	\def\d{1}
	
	
	\node [label = 215 : 0, label = 45 : 3] at (-\d, 2 * \d) {};
	\node [label = 215 : 1, label = 45 : 6] (node-16) at (0, 2 * \d) {};
	\node [label = 215 : 2, label = 45 : 5] (node-25) at (\d, 2 * \d) {};
	\node [label = 215 : 4, label = 45 : 7] (node-47) at (-\d, \d) {};
	
	\draw [rounded corners] (node-16.west) -- (-2 * \d, 2 * \d) -- (-2 * \d, 1.75 * \d);
	
	\begin{scope}[decoration = {markings, mark = at position 0.5 with {\arrow{latex}}}]
	
		\draw [postaction = {decorate}] (node-16.east) -- (node-25.west);
		\draw [postaction = {decorate}, rounded corners] (node-25.south) -- (\d, 2.75 * \d) -- (0, 2.75 * \d) -- (node-16.south);
		
		\draw [postaction = {decorate}] (-2 * \d, 1.25 * \d) -- (-2 * \d, 1.75 * \d);
		\draw [postaction = {decorate}] (-0.5 * \d, 0.5 * \d) -- (0.5 * \d, 0.5 * \d);
		\draw [postaction = {decorate}] (\d, 0.75 * \d) -- (node-25.south);
		\draw [postaction = {decorate}] (1.5 * \d, 3 * \d) -- (-0.5 * \d, 3 * \d);
	
	\end{scope}
	
	\begin{scope}[rounded corners]
	
		\draw [draw = white, double = black, double distance between line centers = 3 pt, line width = 2.6 pt] (-0.5 * \d, 3 * \d) -- (-\d, 3 * \d) -- (-\d, 0.5 * \d) -- (-0.5 * \d, 0.5 * \d);
	
		\draw (-2 * \d, 1.25 * \d) -- (-2 * \d, \d) -- (node-47.west);
		\draw (node-47.east) -- (0, \d) -- (node-16.south);
	
		\draw (node-25.east) -- (2 * \d, 2 * \d) -- (2 * \d, 3 * \d) -- (1.5 * \d, 3 * \d);
		\draw (0.5 * \d, 0.5 * \d) -- (\d, 0.5 * \d) -- (\d, 0.75 * \d);
	
	\end{scope}

	\draw (-\d, 2 * \d) circle (0.15);
	\draw (0, 2 * \d) circle (0.15);

\end{tikzpicture}
\caption{\label{DT-seq-ex}A visual representation of the DT sequence $3^\ell 5^- 7^+ 1^u$.}
\end{figure}
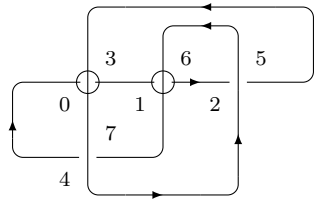

\begin{enumerate}

	\item For a graph with $N$ nodes, the sequence is made of pairs of node labels drawn from the set $\{0, \cdots, 2N - 1\}$.
	
	\item The abbreviated form of the DT sequence will be a listing of odd numbers, given in order of their even counterpart labels in each pair.
	
	\item A crossing is a negative crossing type if the edge with the even label passes under the edge with the odd label, and a positive crossing has the reverse; a number representing a crossing with have either a $+$ or $-$ superscript to indicate the crossing type.

	\item In a similar manner, the vertices are denoted by $\ell$ or $u$, based on the crossing in the knot diagram corresponding to the vertex. If the even edge passes under the odd label in the knot diagram, the vertex is labeled as $\ell$; otherwise, the vertex is labeled $u$.

\end{enumerate}
An example is shown in Figure \ref{DT-seq-ex} of the graph sequence $3^\ell 5^- 7^+ 1^u$. As mentioned above, a given knot has many equivalent DT sequences; the same is true for knotted graphs. Thus, we shall use the sequence that comes lowest in lexicographic order, where the superscripts are taken from lowest to greatest order as $\ell, -, +, u$. Again, as with knots, this only counts transformations of the node labels, and does not include the graph equivalence moves changing one graph diagram to another.

Here, we make one final comment regarding DT sequences of a graph. To enumerate graph sequences, we can use an algorithm similar to what Dowker and Thistletwaite used for knots~\cite{Dow-Thi83}. That algorithm searches for all sequences of $2N$ integers that correspond to a closed path, where the only crossings are those associated with the pairs of numbers in the sequence. In the case of knotted 4-regular graphs, this would list all possible graphs of $N$ nodes, not vertices. Thus, as $N$ increases, graphs with less than $N$ vertices will be included in the list, where the remaining nodes are crossings. This leads to the question of what is the smallest number of nodes necessary to describe a given graph. We denote this as the {\it node number} $n(G)$ of the graph $G$, and is analogous to the crossing number of a knot. When enumerating all graphs, starting from $N = 1$ and incrementing, it is obvious what the node number of a given graph is -- if the graph did not appear for an earlier node number, then it is described by the lowest number where it does appear. However, for a generic graph, the node number of a graph is not obvious; this will be an issue when considering graphs obtained by the Pachner moves in Section \ref{Pachner}. The discussion earlier in this section, on a step-by-step process of making a graph proper, gives a method of finding the DT sequence for a given diagram and choice of Eulerian circuit, but this may not determine the minimal number of nodes necessary to describe the graph. We shall return briefly to this issue in Section \ref{Pachner}.


\section{A two-variable polynomial invariant}
\label{skein-poly}

At this point, we now have sufficient background to start talking about invariant polynomials for knotted 4-regular graphs. However, it will be helpful to spell out something that is implicit in the discussion of Section \ref{notation}. With our choice of the two possible vertex states, there is a natural way to associate the diagram of a knot with a given graph diagram. Specifically, as with Figure \ref{vert-cross} for the $\ominus$ state (and in a like manner for the $\oplus$ state), there is a diagram where the vertex is changed to a crossing of the appropriate type, and the remainder of the graph diagram is the same. By doing this for all vertices, we find the {\it induced knot diagram} matching a given graph diagram with an Eulerian circuit through it. Note for a given graph diagram, each choice of Eulerian circuit through the graph will, in general, give a induced knot diagram for a different knot. When the circuit passes through improper vertices as shown on the left-hand sides of Figures \ref{bad-vert-1} and \ref{bad-vert-2}, then there is no matching crossing in the knot diagram. Thus, the manner in which the circuit passes through the vertices will affect the induced knot diagram for the graph.

\begin{figure}[hbt]
\centering
\begin{tikzpicture}[> = latex, font = \scriptsize]
\matrix[column sep = 0.25 cm]{

	\draw [->] (0, -0.8) node [below] {$D$} -- (0, 0.8) node [above] {$B$};
	\draw [fill = white] (0, 0) circle (0.15);
	\draw [->] (-0.8, 0) node [left] {$A$} -- (0.8, 0) node [right] {$C$};

&
	\draw [->] (0, 0) -- node [above] {RV} (1, 0);
&

	\draw [rounded corners] (-0.8, 0) node [left] {$A$} -- (0.35, 0) -- (0.35, -0.8) node [below] {$D$};
	\draw [fill = white] (0, 0) circle (0.15);
	\draw [<-, rounded corners] (0, 0.8) node [above] {$B$} -- (0, -0.4) -- (0.2, -0.4);
	\draw [draw = white, double = black, double distance between line centers = 3 pt, line width = 2.6 pt] (0.2, -0.4) -- (0.4, -0.4);
	\draw [->] (0.4, -0.4) -- (0.8, -0.4) node [right] {$C$};

	\draw [->, dotted, rounded corners] (0.5, -0.25) -- (0.5, 0.15) -- (0.15, 0.15) -- (0.15, 0.6);
	\draw [->, dotted, rounded corners] (-0.6, -0.15) -- (-0.15, -0.15) -- (-0.15, -0.55) -- (0.15, -0.55);

& 

	\draw [->, rounded corners] (-0.8, 0) node [left] {$A$} -- (0.35, 0) -- (0.35, 0.8) node [above] {$B$};
	\draw [fill = white] (0, 0) circle (0.15);
	\draw [rounded corners] (0, -0.8) node [below] {$D$} -- (0, 0.4) -- (0.2, 0.4);
	\draw [draw = white, double = black, double distance between line centers = 3 pt, line width = 2.6 pt] (0.2, 0.4) -- (0.4, 0.4);
	\draw [->] (0.4, 0.4) -- (0.8, 0.4) node [right] {$C$};

	\draw [->, dotted, rounded corners] (0.15, -0.6) -- (0.15, -0.15) -- (0.5, -0.15) -- (0.5, 0.25);
	\draw [->, dotted, rounded corners] (-0.6, 0.15) -- (-0.15, 0.15) -- (-0.15, 0.55) -- (0.15, 0.55);

&

	\draw [<->, rounded corners] (0.8, 0) node [right] {$C$} -- (-0.35, 0) -- (-0.35, 0.8) node [above] {$B$};
	\draw [fill = white] (0, 0) circle (0.15);
	\draw [rounded corners] (0, -0.8) node [below] {$D$} -- (0, 0.4) -- (-0.2, 0.4);
	\draw [draw = white, double = black, double distance between line centers = 3 pt, line width = 2.6 pt] (-0.2, 0.4) -- (-0.4, 0.4);
	\draw (-0.4, 0.4) -- (-0.8, 0.4) node [left] {$A$};
	
	\draw [->, dotted, rounded corners] (-0.15, -0.6) -- (-0.15, -0.15) -- (-0.5, -0.15) -- (-0.5, 0.25);
	\draw [<-, dotted, rounded corners] (0.6, 0.15) -- (0.15, 0.15) -- (0.15, 0.55) -- (-0.15, 0.55);

&

	\draw [<-, rounded corners] (0.8, 0) node [right] {$C$} -- (-0.35, 0) -- (-0.35, -0.8) node [below] {$D$};
	\draw [fill = white] (0, 0) circle (0.15);
	\draw [<-, rounded corners] (0, 0.8) node [above] {$B$} -- (0, -0.4) -- (-0.2, -0.4);
	\draw [draw = white, double = black, double distance between line centers = 3 pt, line width = 2.6 pt] (-0.2, -0.4) -- (-0.4, -0.4);
	\draw (-0.4, -0.4) -- (-0.8, -0.4) node [left] {$A$};

	\draw [->, dotted, rounded corners] (-0.5, -0.25) -- (-0.5, 0.15) -- (-0.15, 0.15) -- (-0.15, 0.6);
	\draw [<-, dotted, rounded corners] (0.6, -0.15) -- (0.15, -0.15) -- (0.15, -0.55) -- (-0.15, -0.55);

\\
};
\end{tikzpicture}
\caption{\label{orient-rot}The effect of the RV vertex rotation move on a vertex, with a choice of Eulerian circuit through the vertex. The arrows on the edges indicate the orientations of the outgoing edges. The original vertex on the left has the Eulerian cycle passing left to right, and bottom to top, through the vertex. The four right-hand vertices are the possible results of an RV move on the vertex. For each of these cases, the dotted arrows show how the Eulerian cycle now passes around the vertex after the move.}
\end{figure}
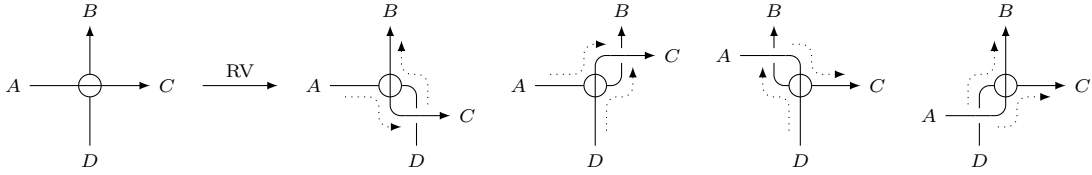

The knot corresponding to the induced knot diagram, associated with a choice of graph diagram and its Eulerian circuit, is invariant under the graph equivalence moves. This is obviously true for the first three Reidemeister moves, since they are taken directly from knot theory. The RIV move for the graph diagram simply moves the vertex from one side of an edge to the other; for the induced knot diagram, this is the same as an RIII move. Finally, we look at the RV vertex rotation move. It is here that the requirement of diffeomorphism invariance for the graph embedding is important, since it restricts the possible vertex rotations to only those that preserve invariance of the induced knot diagram. To see why this is, we consider the case shown in Figure \ref{orient-rot}, starting with the vertex on the left-hand side. The Eulerian circuit is chosen so that it moves through the vertex $A \to C$ and $D \to B$. All allowed results for an RV move acting on this vertex are shown on the right-hand side of the same figure. Only four final situations occur; the others do not arise because of the starting vertex state. Notice how the induced knot diagrams are affected by the vertex rotation. Originally, the induced knot diagram would have a crossing at the location of the vertex. After the RV move, the Eulerian circuit passes around the vertex from the ingoing edges to the outgoing edges as shown by the dotted arrows. None of the four results have the circuit crossing over itself at the vertex. Instead, there is now a crossing on two of the edges incident to the vertex. In particular, the crossings are such that the final induced knot diagram is equivalent to the original -- in all cases, the $A \to C$ portion passes over the $D \to B$ part. This is the key fact that allows polynomial invariants of the graph diagram to be defined. Althought it is not shown here for all cases, this is true for any arrangement of vertex state, edge orientations, and connections between ingoing and outgoing edges (so the vertex does not have to be in proper form, as in this example).

With this discussion in mind, we now define a two-variable polynomial on graph diagrams, for a given choice of Eulerian circuit. The basic relations for this invariant are given in equations (\ref{poly-inv}). These are defined by their action on each vertex of the graph diagram. For each relation in (\ref{poly-inv}), only the portion of the graph diagram shown will be changed; the rest of the diagram remains the same. The left-hand side of the relations give the input vertex, and how the Eulerian circuit passes through this chosen vertex (thus, the brackets have a subscript $G$ for ``graph"). On the right, the vertex is replaced in the diagram by the given new connection of the edges. This new connection is evaluated for any choice of invariant for knots and links (so the brackets have subscript $L$). For our purposes, the specific knot invariant used on the right-hand sides of (\ref{poly-inv}) does not matter, as long as it is sufficient to distinguish the knots and links that arise in the process.

\begin{subequations}
\label{poly-inv}
\begin{eqnarray}
\label{prop-rel}
	\propervertex &=& c \propercross + s \propersplit	\\
\label{improp-rel}
	\impropervertex &=& c \propersplit + s \propercross		\\
\label{rev-rel}
	\rvertex &=& c \nesplit + s \nwsplit
\end{eqnarray}
\end{subequations}
All other possibilities can be obtained from these by reflections of the given portions of the graph diagram, or reversing the orientation of either the vertical or horizontal edges, or both. Note that the relations on the first two lines depend on the vertex state; reversing this will lead to the opposite type of crossing on the right-hand side of each relation. However, the vertex state in the third relation is irrelevant to the vertex-less states on the right.

Checking that these relations (\ref{poly-inv}) are invariant under the Reidemeister moves is relatively straightforward. The first three moves, RI-RIII, do not affect vertices at all. As mentioned above, the RIV move on the graph diagram would correspond to an RIII move on the knot diagram resulting from these relations. Finally, for the RV graph rotation, note that each such rotation preserves the number and type of crossing in the induced knot diagram of the graph. In the $c$ terms of the relations (\ref{poly-inv}), the type of crossing in the induced knot diagram is preserved, and so is maintained under the RV move. For the $s$ terms, on the other hand, the only other possible connection of the oriented edges is used. This is done in such a way that the crossing that results from the graph rotation move is the opposite of that for the $s$ terms in the second and third relations (\ref{improp-rel}) and (\ref{rev-rel}), so it can be eliminated by a RI move. Thus, the lack of crossing in the (\ref{prop-rel}) $s$ term remains the same in the knot diagram. These considerations show that the polynomial obtained by these relations is invariant under all the Reidemeister moves.
\begin{proposition}
	Relations (\ref{poly-inv}) acting on a graph diagram, with a particular choice of Eulerian circuit and invariant $[\ ]_L$ of links, define a two-variable polynomial for the graph. This polynomial is invariant under the graph equivalence moves.
\end{proposition}

In addition, it is possible to evaluate these relations on all possible Eulerian circuits for a given graph diagram. This will result in a multiset of invariant polynomials associated with the graph, similar in spirit to the set $C(G)$ defined by Kauffman~\cite{Kau89}, a collection of link diagrams associated with a single graph diagram. Note that this gives information in two different ways. First, the number of multiset members is the same as the number of distinct Eulerian circuits for the graph. However, the actual members of the multiset help to distinguish, e.g. two different embeddings of the same abstract graph (which would equal numbers of circuits). If there is a polynomial that is present in one multiset, but not the other, for example, this is sufficient to distinguish the two graphs.

\begin{corollary}
	Let $S_G$ be the multiset of polynomials obtained by evaluating the relations (\ref{poly-inv}) for all possible Eulerian circuits for a graph diagram $G$. The set $S_G$ is invariant under the graph equivalence moves.
\end{corollary}

Table \ref{graph-list} shows the properties of all graphs that have proper graph diagrams of three nodes or less. These properties are described in terminology borrowed from knot theory. First, a graph diagram is {\it chiral} when it is not equivalent to its mirror image. In other words, there is no possible sequence of Reidemeister moves that will change a chiral graph diagram into its mirror version. For graphs that are amphicheiral (or achiral), on the other hand, there is such a sequence. An example of such a sequence has been found for all entries in Table \ref{graph-list} given as amphicheiral. Thus, for the chiral graphs in Table \ref{graph-list}, the DT sequence represents two separate graphs, rather than just a single graph. Next, when we include both the even and odd labels, a DT sequence can be thought of as a map $a$ from all integers $\{0, 1, \cdots, 2N - 1\}$ to itself. A DT sequence is called {\it prime} if no proper subinterval $\{k, k + 1, \cdots, l\}$ (modulo $2N$) of this set of numbers is mapped to itself under $a$. Thus, for example, the DT sequence $1^\ell 3^\ell$ is not prime, since $a$ maps $\{0, 1\}$ to itself, while the sequence $3^\ell 5^- 1^-$ is prime, since $a$ maps no continuous proper sequence of numbers to itself. The origin of this terminology is from knot theory, where a prime knot cannot be written as the connected sum of two other knots. Much like prime numbers, prime knots are sufficient to construct all knots. However, prime graphs do not have exactly the same property. The location of the connected sum for knots does not matter, since there is freedom to move one of the knots in the sum to any location of the second knot. The vertices of a graph prevent this from happening, so connected sums of the same two graph diagrams can give differing results. However, the condition of prime graphs as defined here in terms of DT sequences still remains, and may serve as a useful property for graph classification in the future.

\begin{table}[hbt]
\centering
\begin{tabular}{|c|c|c||c|c|c|}
\hline
Graph sequence		& Prime?	& Chiral?	& Graph sequence		& Prime?	& Chiral?	\\
\hline
$1^\ell$			& Y		& N 		& $3^\ell 5^- 1^u$		& Y		& N 		\\
$1^\ell 3^\ell$		& N 		& N 		& $3^\ell 5^\ell 1^\ell$	& Y		& Y		\\
$3^\ell 5^- 1^-$	& Y 		& Y		& $3^\ell 5^\ell 1^u$		& Y		& Y		\\
$3^\ell 5^+ 1^+$	& Y		& N 		& $1^\ell 3^\ell 5^\ell$	& N 		& N 		\\
$3^\ell 5^\ell 1^-$	& Y		& Y		& $1^\ell 5^\ell 3^\ell$ 	& N 		& N 		\\
$3^\ell 5^\ell 1^+$	& Y		& N 		&					&		&		\\
\hline
\end{tabular}
\caption{\label{graph-list}List of all knotted 4-regular knotted graphs with diagrams of three or less nodes. For each graph, its DT sequence is given, along with whether the graph is prime and chiral.}
\end{table}

\begin{figure}[hbt]
\centering
\begin{tikzpicture}[> = latex]
\matrix[column sep = 1 cm]{

	
	\begin{scope}[decoration = {markings, mark = at position 0.5 with {\arrow{latex}}}]

		\draw [postaction = {decorate}] (1, 0.75) -- (-0.75, 0.75);
		\draw [postaction = {decorate}] (-0.75, -0.5) -- (-0.25, -0.5);
		\draw [postaction = {decorate}] (0.25, 0.5) -- (0.75, 0.5);
		\draw [postaction = {decorate}] (0.75, -0.75) -- (-0.75, -0.75);

		\draw [postaction = {decorate}] (-0.85, 0) -- (-0.15, 0);
		\draw [postaction = {decorate}] (0.15, 0) -- (0.85, 0);
	
	\end{scope}
	
	\begin{scope}[rounded corners]
	
		\draw (1.15, 0) -- (1.5, 0) -- (1.5, 0.75) -- (1, 0.75);
		\draw (-0.75, 0.75) -- (-1, 0.75) -- (-1, -0.5) -- (-0.75, -0.5);
		\draw (-0.25, -0.5) -- (0, -0.5) -- (0, 0.5) -- (0.25, 0.5);
		\draw (0.75, 0.5) -- (1, 0.5) -- (1, -0.75) -- (0.75, -0.75);
		\draw (-0.75, -0.75) -- (-1.5, -0.75) -- (-1.5, 0) -- (-1.15, 0);

	\end{scope}


	\draw (-1, 0) circle (0.15);
	\draw [fill = white] (0, 0) circle (0.15);
	\draw (1, 0) circle (0.15);


	\draw (-0.15, 0) -- (0.15, 0);


	\node at (0, -1.5) {$3^\ell 5^\ell 1^\ell$};

&


	\draw (-1, 0) circle (0.15);
	\draw (0, 0) circle (0.15);
	\draw (1, 0) circle (0.15);

	
	\begin{scope}[decoration = {markings, mark = at position 0.5 with {\arrow{latex}}}]

		\draw [postaction = {decorate}] (1, 0.75) -- (-0.75, 0.75);
		\draw [postaction = {decorate}] (-0.75, -0.5) -- (-0.25, -0.5);
		\draw [postaction = {decorate}] (0.25, 0.5) -- (0.75, 0.5);
		\draw [postaction = {decorate}] (0.75, -0.75) -- (-0.75, -0.75);

		\draw [postaction = {decorate}] (-0.85, 0) -- (-0.15, 0);
		\draw [postaction = {decorate}] (0.15, 0) -- (0.85, 0);
	
	\end{scope}
	
	\begin{scope}[rounded corners]
	
		\draw (1.15, 0) -- (1.5, 0) -- (1.5, 0.75) -- (1, 0.75);
		\draw (-0.75, 0.75) -- (-1, 0.75) -- (-1, -0.5) -- (-0.75, -0.5);
		\draw (-0.25, -0.5) -- (0, -0.5) -- (0, 0.5) -- (0.25, 0.5);
		\draw (0.75, 0.5) -- (1, 0.5) -- (1, -0.75) -- (0.75, -0.75);
		\draw (-0.75, -0.75) -- (-1.5, -0.75) -- (-1.5, 0) -- (-1.15, 0);

	\end{scope}


	\begin{scope}[->, dotted, rounded corners]
	
		\draw (-0.85, 0.6) -- (-0.85, 0.15) -- (-0.3, 0.15);
		\draw (-1.4, -0.5) -- (-1.4, -0.15) -- (-1.1, -0.15) -- (-1.1, -0.5);
		
		\draw (0.3, -0.15) -- (0.85, -0.15) -- (0.85, -0.6);
		\draw (1.1, 0.5) -- (1.1, 0.15) -- (1.4, 0.15) -- (1.4, 0.5);
	
	\end{scope}


	\node at (0, -1.5) {$3^\ell 5^\ell 1^u$};

\\
};
\end{tikzpicture}
\caption{\label{3-vert-diag}Graph diagrams for the sequences $3^\ell 5^\ell 1^\ell$ and $3^\ell 5^\ell 1^u$. On the left side, the Eulerian circuit shown for $3^\ell 5^\ell 1^\ell$ (and its globally reversed version) is the only one that gives a trefoil knot for the coefficient of $c^3$, after acting with the relations (\ref{poly-inv}). The right side shows the only Eulerian circuit (up to reversal) which gives an oriented Hopf link for the coefficient of $s^3$.}
\end{figure}
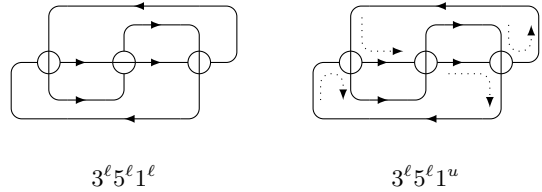

As an example of how the results in Table \ref{graph-list} are obtained, we consider the two sequences $3^\ell 5^\ell 1^\ell$ and $3^\ell 5^\ell 1^u$. Graph diagrams for these sequences are shown in Figure \ref{3-vert-diag}, along with example orientations. These sequences are the only two prime DT sequences with three vertices (up to the label shifting mentioned in Section \ref{notation}). Thus, it is useful to show that they represent distinct graphs. We will also show that they are both chiral. We start with $3^\ell 5^\ell 1^\ell$. Since the base sequence 3 5 1 corresponds to the knot DT sequence 4 6 2 of the trefoil knot, it is possible that the trefoil knot appears as the induced knot diagram of the sequence $3^\ell 5^\ell 1^\ell$. This could occur only if all three vertices are proper (as shown in the left-hand Figure \ref{3-vert-diag}), or else all three are improper vertices of the form shown in Figure \ref{bad-vert-2}. However, the latter case would not give a single Eulerian circuit for the entire graph diagram. Thus, the trefoil knot appears only twice -- for the orientation shown in Figure \ref{3-vert-diag} and its reversed version -- as the coefficient of $c^3$ in the multiset of invariant polynomials. Now we use the fact that the trefoil knot is chiral. Reversing the orientation of a knot does not affect its chirality, so both orientations give the same version of the trefoil knot. Taking the mirror image of the graph diagram shown in Figure \ref{3-vert-diag} would give a $c^3$ coefficient with the opposite chirality. Therefore, this induced knot is distinct from its mirror image, implying the graph $3^\ell 5^\ell 1^\ell$ is chiral as well. Turning to the other three-vertex graph $3^\ell 5^\ell1 ^u$, it is not possible to have a trefoil knot appear in any of the coefficients of the invariant polynomial, because of the new choice of vertex states. Note that the relations (\ref{poly-inv}) do not change the crossing types of induced knot diagrams. However, it is possible to get the oriented Hopf link, comprised of two interlocked circles. In particular, with the graph orientation shown in Figure \ref{3-vert-diag} for $3^\ell 5^\ell 1^u$, this link appears as the coefficient of $s^3$ in the associated invariant polynomial. Up to orientation reversal, this orientation is the only possiblity for the Hopf link to appear as the $s^3$ coefficient. Since the oriented Hopf link is chiral, then using the same logic as above, the graph $3^\ell 5^\ell 1^u$ is as well.

As a final comment, the two-variable polynomial described in this section was derived by the expedient of trying the simplest possibilities for relations acting on the graph vertices, inspired by similar relations for the Jones and HOMFLY-PT polynomials. However, these latter knot polynomials are part of a family of quantum invariants, based on a choice of quantum group. It may be fruitful to consider how graph polynomial invariants, such as the one presented here, arise in a like fashion. Such invariants may have a physical interpretation coming from the action defined over the space of 4-valent graphs. This is because quantum groups naturally arise when the cosmological constant is considered in LQG; in such an approach, the graphs are framed, where the edges and vertices are thickened to become tubes and spheres~\cite{Bor-Maj-Smo96}. This fits in naturally with the process of going from the graph diagram to its accompanying induced knot diagram.

\section{A quandle counting polynomial invariant}
\label{quandle-poly}

\begin{figure}[hbt]
\centering
\begin{tikzpicture}[> = latex]

	\draw (-0.5, 0) node [left] {$a \triangleleft b$} -- (-0.15, 0)
		(0.15, 0) -- (0.5, 0) node [right] {$a$};
	\draw [->] (0, -0.5) node [below] {$b$} -- (0, 0.5) node [above] {$b$};

\end{tikzpicture}
\caption{\label{quandle-cross}Coloring of the arcs incident to a crossing in an oriented knot or link. Only the orientation of the top arc is relevant to the coloring.} 
\end{figure}
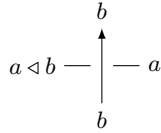

The second polynomial invariant we define is based on the idea of a knot quandle. For completeness, we first give a short summary of the definition of a quandle; a more in-depth review can be found with the monograph by Elhamdadi and Nelson~\cite{Elh-Nel15}. Quandles are algebraic objects used to distinguish knots and links. For example, a simple invariant is the number of ways to color the arcs of a knot diagram by elements of a quandle. The rules governing a quandle are chosen so that this number remains the same under the first three Reidemeister moves. Since the notation for a quandle varies quite a bit in the literature, the definition of this algebraic object is given completely. A {\it quandle} $Q$ is a set with a binary operation $\triangleleft$ defined on it such that the following properties are true, for $a, b, c \in Q$.
\begin{enumerate}

	\item $a \triangleleft a = a$;
	\item \label{Q-inverse}for every pair $a, b$ there is a unique $c$ such that $a = c \triangleleft b$; and
	\item \label{Q-dist}$(a \triangleleft b) \triangleleft c = (a \triangleleft b) \triangleleft (b \triangleleft c)$

\end{enumerate}
Each of these conditions on the elements of the quandle arises from the need to preserve the edge colorings for the rest of the graph under one of the first three Reidemeister moves (in fact, the move with the same number as given in the list above). Note that condition (\ref{Q-inverse}) implies that there is an inverse operation $\widetilde{\triangleleft}$, such that if $a = c \triangleleft b$, then $c = a \tildetriangleleft b$. An involutive quandle (also known as a {\it kei}) has $(a \triangleleft b) \triangleleft b = a$, so that $\triangleleft$ and $\widetilde{\triangleleft}$ are the same operation. Figure \ref{quandle-cross} shows how the arcs incident to a crossing are colored by elements of a quandle. For the edge passing over the crossing, the quandle element $b$ is the same on both sides of the crossing. The remaining two arcs have a coloring dependent on how they are seen by an observer moving along the orientation of the top edge. When an element $a$ is on the arc incident to the right of the crossing (as seen by the overcrossing edge), the arc on the left must be colored with the element $a \triangleleft b$.

\begin{figure}[hbt]
\centering
\begin{tikzpicture}[> = latex]
\matrix[column sep = 0.5 cm]{


	\draw (2.5, 0) node [left] {$a \triangleleft b$} -- (3.5, 0) node [right] {$a$};
	\draw [fill = white] (3, 0) circle (0.15);
	\draw [->] (3, -0.5) node [below] {$b$} -- (3, 0.5) node [above] {$b$};

&


	\draw [->] (-0.5, 0) node [left] {$a$} -- (0.5, 0) node [right] {$b$};
	\draw [->] (0, -0.5) node [below] {$b$} -- (0, 0.5) node [above] {$a$};
	\draw [fill = gray!50] (0, 0) circle (0.15);
	
	\draw [->, dotted, rounded corners] (0.15, -0.5) -- (0.15, -0.15) -- (0.5, -0.15);
	\draw [->, dotted, rounded corners] (-0.5, 0.15) -- (-0.15, 0.15) -- (-0.15, 0.5);

&


	\draw (-0.5, 0) node [left] {$a$} -- (0.5, 0) node [right] {$b$};
	\draw [<->] (0, -0.5) node [below] {$a$} -- (0, 0.5) node [above] {$b$};
	\draw [fill = gray!50] (0, 0) circle (0.15);

	
	\draw [->, dotted, rounded corners] (0.5, 0.15) -- (0.15, 0.15) -- (0.15, 0.5);
	\draw [->, dotted, rounded corners] (-0.5, -0.15) -- (-0.15, -0.15) -- (-0.15, -0.5);

\\
};

\end{tikzpicture}
\caption{\label{quandle-vert}Quandle colorings of edges incident to a vertex in a graph diagram. The basic rule is to color the edges respecting the relations in the induced knot diagram corresponding to the graph. Thus, the quandle elements are the same for the last two situations, regardless of the vertex state.}
\end{figure}
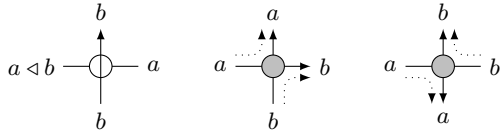

Our strategy for developing a polynomial invariant based on quandles is to use the induced knot diagrams associated with the graph. For each Eulerian circuit of the graph, one can count the number of colorings of its arcs by quandle elements. How this is done at a vertex is shown in Figure \ref{quandle-vert}. Specifically, a proper vertex is treated exactly as the corresponding crossing in the induced knot diagram. The arc that crosses ``over" the vertex is colored with a single quandle element $b$, while the other two arcs are colored $b$ or $a \triangleleft b$, depending on whether they are to the right or left, respectively, of the oriented edge passing ``over" the vertex. For improper vertices, there is no crossing in the induced knot diagram. Thus, each pair of an ingoing edge, and its subsequent outgoing edge in the Eulerian circuit, are colored by the same quandle element. These latter two arc colorings do not depend on the vertex state.

With this scheme in mind, the arcs of a graph diagram, with the arc orientations given by an Eulerian circuit, can be colored by the elements of a chosen quandle. This results in a coloring of the induced knot diagram; this coloring is invariant under the Reidemeister moves, leading to the following result.

\begin{proposition}
	Let $D$ be a graph diagram of a knotted 4-regular pseudograph, whose arcs are oriented by a choice of Eulerian circuit $E$ on the diagram, and $Q$ be a quandle. The number of colorings $|C_{Q, E} (D)|$ of the arcs of $D$ by elements of $Q$, obeying the conditions given by Figure \ref{quandle-vert} for arcs incident to a vertex, is an invariant under the graph equivalence moves.
\end{proposition}
However, we can go one step further than this, and define a quandle coloring polynomial that contains more information. Fixing a graph diagram $D$ and a quandle $Q$, we can find the number of colorings $|C_{Q, E}|$ for every choice $E$ of Eulerian circuit on the graph. This will give a multiset $S_C$ of coloring numbers. Let $\Psi_Q$ be the generating function of this multiset, in an abstract variable $u$. Thus, each term $n u^c$ will denote the fact that the number $c$ appears in the multiset $S_C$ $n$ times, i.e. there are $n$ Eulerian circuits $E$ such that the number of graph colorings $|C_{Q, E} (D)|$ is $c$. Since each coloring number is invariant, then so will this polynomial $\Psi$.

\begin{corollary}
	The generating function $\Psi_Q (G)$ of the number of colorings $|C_{Q, E}(D)|$, for a choice of quandle $Q$, any graph diagram $D$ of the graph $G$, and all possible Eulerian circuits $E$ of the graph, is invariant under the graph equivalence moves.
\end{corollary}
Note that $\Psi_Q (u)$ evaluated at $u = 1$ is simply the number of Eulerian circuits of the graph; in addition, the set of powers in the polynomial is an invariant, since it is the set of the number of colorings of the graph diagram.

\begin{figure}[hbt]
\centering
\begin{tikzpicture}[> = latex]


	\draw [draw = white, double = black, double distance between line centers = 3 pt, line width = 2.6 pt, rounded corners](1.5, 0.5) -- (1.5, 0.75) -- (-1, 0.75) -- (-1, -0.5) -- (-0.5, -0.5);
	\draw [draw = white, double = black, double distance between line centers = 3 pt, line width = 2.6 pt, rounded corners] (-1.5, -0.5) -- (-1.5, -0.75) -- (1, -0.75) -- (1, 0.5) -- (0.5, 0.5);
	\draw [draw = white, double = black, double distance between line centers = 3 pt, line width = 2.6 pt, rounded corners] (-1.5, -0.5) -- (-1.5, 0) -- (1.5, 0) -- (1.5, 0.5);
	\draw [draw = white, double = black, double distance between line centers = 3 pt, line width = 2.6 pt, rounded corners] (-0.5, -0.5) -- (0, -0.5) -- (0, 0.5) -- (0.5, 0.5);


	\node [left] at (-1.5, -0.375) {$a$};
	\node [right] at (1.5, 0.375) {$b$};
	\node [above] at (-0.5, -0.5) {$c$}; 

\end{tikzpicture}
\caption{\label{trefoil}The trefoil knot, colored by three quandle elements $a, b, c$.}
\end{figure}
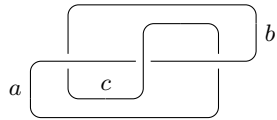

As an example, we return to the two graph diagrams $3^\ell 5^\ell 1^\ell$ and $3^\ell 5^\ell 1^u$ shown in Figure \ref{3-vert-diag}, and compute their quandle counting polynomials. Since the only difference between the two diagrams are the vertex states, then both have the same number of Eulerian circuits -- in this case, 32. However, unlike the two-variable polynomial of Section \ref{skein-poly}, there is only one way to color the edges incident to each vertex, based on how the Eulerian circuit passes through that vertex. In other words, only the knots coming from the $c$ terms of the polynomial from Section \ref{skein-poly} appear here, not any of the $s$ terms. Since the sequence $3^\ell 5^\ell 1^\ell$ had the trefoil appear as the coefficient of $c^3$, then there are two circuits that give it here as well (the circuit shown to the left of Figure \ref{3-vert-diag}, and its reversed version). All of the remaining circuits will simply give some version of the unknot. Now, we choose a quandle to work with. For our purposes, it is sufficient to use the Takasaki kei $\mathbb{Z}_3$, whose operation table is shown in Table \ref{kei}.
\begin{table}[hbt]
\centering
\begin{tabular}{c|ccc}
$\triangleleft$	& 0 & 1 & 2 \\
\hline
0			& 0 & 2 & 1 \\
1			& 2 & 1 & 0 \\
2			& 1 & 0 & 2 \\
\end{tabular}
\caption{\label{kei}The Takasaki kei $\mathbb{Z}_3$, where the quandle operation is given by $x \triangleleft y = 2y - x$ for $x, y \in \mathbb{Z}_3$.}
\end{table}
This operation $\triangleleft$ on two elements $x, y \in \mathbb{Z}_3$ is given by $x \triangleleft y = 2y - x$, where $x$ is the row of the table, and $y$ the column. For this kei, the unknot has three colorings total, where every arc in the induced knot diagram is colored by the same kei element. The trefoil knot, however, has a different number of colorings. An example of this knot is given in Figure \ref{trefoil}; the quandle elements $a, b, c$ labeling the arcs as in this diagram must satisfy
\[
	a = b \triangleleft c \qquad
	b = c \triangleleft a \qquad
	c = a \triangleleft b
\]
These relations are independent of the knot orientation. With this choice of quandle and arc elements, there are again three trivial colorings, and then six additional colorings, with each arc colored by a different element. Returning to $3^\ell 5^\ell 1^\ell$, since there are 30 Eulerian circuits that give the unknot, and two that give the trefoil, its quandle coloring invariant polynomial is
\[
	\Psi_{\mathbb{Z}_3} (3^\ell 5^\ell 1^\ell) = 30u^3 + 2u^9
\]
On the other hand, none of the circuits for the sequence $3^\ell 5^\ell 1^u$ will give the trefoil, since it only appeared in the $cs^2$ term of the two-variable polynomial. Said in a different way, there is no way to get a crossing in the induced knot diagram that is opposite to the vertex state. Thus, the quandle polynomial for this sequence is
\[
	\Psi_{\mathbb{Z}_3} (3^\ell 5^\ell 1^u) = 32u^3
\]
with all Eulerian circuits giving the unknot.

\section{Pachner moves}
\label{Pachner}

\subsection{Evolution moves on the space of graphs}
\label{evo-moves}

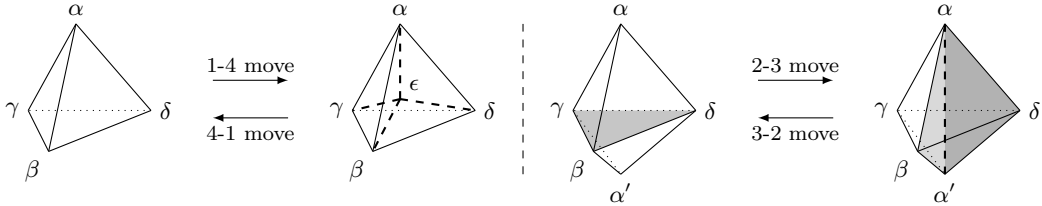
\begin{figure}[hbt]
\centering
\begin{tikzpicture}[> = latex]


\matrix[column sep = 0.25 cm]{

	\coordinate (v1) at (0, -1/3, {sqrt(8/9)});
	\coordinate (v2) at ({sqrt(2/3)}, -1/3, {-sqrt(2/9)});
	\coordinate (v3) at ({-sqrt(2/3)}, -1/3, {-sqrt(2/9)});
	\coordinate (v4) at (0, 1, 0);

	\draw (v1) node [below left] {$\beta$} -- (v2) node [right] {$\delta$};
	\draw (v1) -- (v3) node [left] {$\gamma$};
	\draw (v1) -- (v4) node [above] {$\alpha$};
	\draw [dotted] (v2) -- (v3);
	\draw (v2) -- (v4);
	\draw (v3) -- (v4);

&

	\begin{scope}[->, font = \footnotesize]

		\draw (0, 0.25) -- node [above] {1-4 move} (1, 0.25);
		\draw (1, -0.25) -- node [below] {4-1 move} (0, -0.25);

	\end{scope}

&

	\coordinate (v0) at (0, 0, 0);
	\coordinate (v1) at (0, -1/3, {sqrt(8/9)});
	\coordinate (v2) at ({sqrt(2/3)}, -1/3, {-sqrt(2/9)});
	\coordinate (v3) at ({-sqrt(2/3)}, -1/3, {-sqrt(2/9)});
	\coordinate (v4) at (0, 1, 0);

	\draw (v1) node [below left] {$\beta$} -- (v2) node [right] {$\delta$};
	\draw (v1) -- (v3) node [left] {$\gamma$};
	\draw (v1) -- (v4) node [above] {$\alpha$};
	\draw [dotted] (v2) -- (v3);
	\draw (v2) -- (v4);
	\draw (v3) -- (v4);

	\begin{scope}[thick, dashed]

		\draw (v0) node [above right] {$\epsilon$} -- (v1);
		\draw (v0) -- (v2);
		\draw (v0) -- (v3);
		\draw (v0) -- (v4);

	\end{scope}

&

	\draw [dashed] (0, -1) -- (0, 1);

&

	\coordinate (v1) at (0, -1/3, {sqrt(8/9)});
	\coordinate (v2) at ({sqrt(2/3)}, -1/3, {-sqrt(2/9)});
	\coordinate (v3) at ({-sqrt(2/3)}, -1/3, {-sqrt(2/9)});
	\coordinate (v4) at (0, 1, 0);
	\coordinate (v4p) at (0, -1, 0);

	\draw [fill = gray!45, draw = none] (v1) -- (v2) -- (v3) -- (v1);

	\draw (v1) node [below left] {$\beta$} -- (v2) node [right] {$\delta$};
	\draw (v1) -- (v3) node [left] {$\gamma$};
	\draw (v1) -- (v4) node [above] {$\alpha$};
	\draw [dotted] (v2) -- (v3);
	\draw (v2) -- (v4);
	\draw (v3) -- (v4);

	\draw (v1) -- (v4p) node [below] {$\alpha'$};
	\draw (v2) -- (v4p);
	\draw [dotted] (v3) -- (v4p);

&

	\begin{scope}[->, font = \footnotesize]

		\draw (0, 0.25) -- node [above] {2-3 move} (1, 0.25);
		\draw (1, -0.25) -- node [below] {3-2 move} (0, -0.25);

	\end{scope}

&

	\coordinate (v1) at (0, -1/3, {sqrt(8/9)});
	\coordinate (v2) at ({sqrt(2/3)}, -1/3, {-sqrt(2/9)});
	\coordinate (v3) at ({-sqrt(2/3)}, -1/3, {-sqrt(2/9)});
	\coordinate (v4) at (0, 1, 0);
	\coordinate (v4p) at (0, -1, 0);

	\draw [fill = gray!30, draw = none] (v4) -- (v1) -- (v4p) -- (v4);
	\draw [fill = gray!60, draw = none] (v4) -- (v2) -- (v4p) -- (v4);
	\draw [thick, dashed] (v4) -- (v4p);

	\draw (v1) node [below left] {$\beta$} -- (v2) node [right] {$\delta$};
	\draw (v1) -- (v3) node [left] {$\gamma$};
	\draw (v1) -- (v4) node [above] {$\alpha$};
	\draw [dotted] (v2) -- (v3);
	\draw (v2) -- (v4);
	\draw (v3) -- (v4);

	\draw (v1) -- (v4p) node [below] {$\alpha'$};
	\draw (v2) -- (v4p);
	\draw [dotted] (v3) -- (v4p);

\\
};

\end{tikzpicture}
\caption{\label{Pach-trig}Pachner moves acting on tetrahedra for the triangulation of a three-dimensional manifold. For the 1-4 move, the vertex $\epsilon$ is added, along with edges $\alpha\epsilon, \beta\epsilon, \gamma\epsilon, \delta\epsilon$. For the 2-3 move, the face $\beta\gamma\delta$ is removed, and the edge $\alpha\alpha'$ is added, along with the faces $\alpha\beta\alpha', \alpha\gamma\alpha',$ and $\alpha\delta\alpha'$.}
\end{figure}

The Pachner moves~\cite{Pac91} (also known as bistellar moves) arise in topology when considering the triangulations of a manifold $M$. For a three-dimensional manifold, a triangulation is a set of 3-simplices (or tetrahedra) whose faces are identified in pairs. Now, consider two triangulations of a three-dimensional $M$. These are related to each other by a finite set of moves modifying a local collection of tetrahedra. The Pachner moves on 3-simplices are the 1-4 move (replacing a single tetrahedron with four, with one new internal vertex), the 2-3 move (replacing two tetrahedra with three, with one new internal edge), and their inverses. These Pachner moves are represented in Figure \ref{Pach-trig}. As discussed in Section \ref{intro}, the moves can be interpreted in terms of 4-regular graphs, by writing them in terms of the dual graph. Here, the interior of each tetrahedron is dual to a graph vertex, and every face of the tetrahedron is dual to a graph edge. With this in mind, there are corresponding Pachner moves for the 4-regular graph. The graph versions of the 1-4 and 2-3 moves are straightforward: a single vertex is replaced by four vertices, or two vertices sharing an edge are replaced by three, respectively. Each of these moves creates 3-cycles in the graph, i.e. triangles of edges between a group of three vertices. Thus, for the inverse 4-1 and 3-2 moves, there is an issue if there is an outside edge passing through these cycles; in this case, it is not clear what the final subgraph would look like. To avoid issues such as these, we assume that the 4-1 and 3-2 moves cannot take place unless there are no such exterior edges. In other words, within the embedding manifold, there is a neighborhood of the subgraph that has the same structure as the result of a 1-4 or 2-3 move~\cite{Smo-Wan08}. With these considerations in mind, the Pachner moves on a 4-regular graph are shown in Figures \ref{1-4} and \ref{2-3} for the 1-4/4-1 and 2-3/3-2 moves, respectively.

The mapping between triangulations of a three-dimensional manifold, and the 4-regular graphs dual to each such triangulation, give a helpful way to characterize the space of knotted 4-regular graphs. In particular, if two graphs $G_1$ and $G_2$ are dual to topologically distinct manifolds $M_1$ and $M_2$, respectively, then there is no finite set of graph Pachner moves from $G_1$ to $G_2$. This is because there would be no analogous sequence of moves from $M_1$ to $M_2$. Thus, it is expected that the space of knotted 4-regular graphs, which we denote as $\cal{G}$, is made of many components, disconnected under equivalence to distinct three-dimensional manifolds. However, this probably does not fully characterize $\cal{G}$. It is likely that there are graphs dual to the same manifold triangulation (at least as abstract graphs), but whose embeddings differ enough so that there are no Pachner move sequences between them. At this point, there is not a good enough characterization of the space $\cal{G}$ of knotted 4-regular graphs to answer this question. However, we will see that the polynomial invariants developed in this work provide a possible means to study the question further.

The graph Pachner moves provide a partial order $\prec$ on $\cal{G}$, where $G_1 \prec G_2$ if $G_2$ is a graph obtained from $G_1$ either by a 1-4 move or a 2-3 move. Without additional information, fleshing out this partial order would require starting from a given graph, and computing all possible graphs obtained by 1-4 and 2-3 moves. This does not supply a satisfactory means of studying the structure of this partial order, since it may require many computational steps to find the sequence of Pachner moves from one graph to another. Thus, it would be better to have a more analytic means of looking at this partial order. With this in mind, we will now look at how quandle counting polynomial $\Psi$ changes under the Pachner moves; much of the analysis here also carries over to other graph invariants developed.

\subsection{The 1-4 move}
\label{1-4-sect}

\begin{figure}[hbt]
\centering
\begin{tikzpicture}[> = latex]
\matrix[column sep = 0.25 cm, row sep = 0.5 cm]{


	\draw (-0.5, 0) node [left] {$A$} -- (0.5, 0) node [right] {$C$};
	\draw [fill = white] (0, 0) circle (0.15);
	\draw (0, -0.5) node [below] {$D$} -- (0, 0.5) node [above] {$B$};

&

	\begin{scope}[->, font = \footnotesize]

		\draw (0, 0.25) -- node [above] {1-4 move} (1, 0.25);
		\draw (1, -0.25) -- node [below] {4-1 move} (0, -0.25);

	\end{scope}

&


	\draw (-1, 0) node [left] {$A$} -- (1, 0) node [right] {$C$};

	\draw [fill = white] (-0.5, 0) circle (0.15);
	\draw [fill = white] (0, 0.5) circle (0.15);
	\draw [fill = white] (0.5, 0) circle (0.15);
	\draw [fill = white] (0, -0.5) circle (0.15);

	\draw [rounded corners] (-0.15, -0.5) -- (-0.5, -0.5) -- (-0.5, 0.5) -- (-0.15, 0.5)
		(0.15, -0.5) -- (0.5, -0.5) -- (0.5, 0.5) -- (0.15, 0.5);

	\draw [draw = white, double = black, double distance between line centers = 3 pt, line width = 2.6 pt] (0, -0.35) -- (0, 0.35);

	\draw (0, 1) node [above] {$B$} -- (0, 0.35);
	\draw (0, -0.35) -- (0, -1) node [below] {$D$};

&
	\draw [dashed] (0, -1) -- (0, 1);
&


	\draw (0, -0.5) node [below] {$D$} -- (0, 0.5) node [above] {$B$};
	\draw [fill = white] (0, 0) circle (0.15);
	\draw (-0.5, 0) node [left] {$A$} -- (0.5, 0) node [right] {$C$};s

&

	\begin{scope}[->, font = \footnotesize]

		\draw (0, 0.25) -- node [above] {1-4 move} (1, 0.25);
		\draw (1, -0.25) -- node [below] {4-1 move} (0, -0.25);

	\end{scope}

&


	\draw (0, 1) node [above] {$B$} -- (0, -1) node [below] {$D$};

	\draw [fill = white] (-0.5, 0) circle (0.15);
	\draw [fill = white] (0, 0.5) circle (0.15);
	\draw [fill = white] (0.5, 0) circle (0.15);
	\draw [fill = white] (0, -0.5) circle (0.15);

	\draw [rounded corners] (-0.5, -0.15) -- (-0.5, -0.5) -- (0.5, -0.5) -- (0.5, -0.15)
		(-0.5, 0.15) -- (-0.5, 0.5) -- (0.5, 0.5) -- (0.5, 0.15);

	\draw [draw = white, double = black, double distance between line centers = 3 pt, line width = 2.6 pt] (-0.35, 0) -- (0.35, 0);

	\draw (-1, 0) node [left] {$A$} -- (-0.35, 0);
	\draw (0.35, 0) -- (1, 0) node [right] {$C$};

\\
};
\end{tikzpicture}
\caption{\label{1-4}1-4 and 4-1 Pachner graph moves.}
\end{figure}
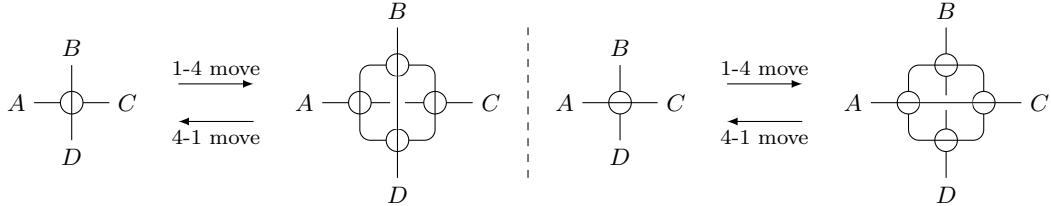
At this point, we consider the effect of the 1-4 move on the graph; the 2-3 move will be examined in Section \ref{2-3-sect}. We shall see that, looking at the knot diagrams induced by the Eulerian circuits on the resulting graph, at least one such knot diagram will be a different knot than the graph before the 1-4 or 2-3 move, although many will be the same as before. The polynomial invariants defined earlier will allow us to quantify this, and therefore also help to flesh out the partial order of the graphs under the 1-4 and 2-3 moves. In Figure \ref{1-4}, the 1-4 move replaces a single vertex with four vertices; all of these new vertices are connected to each other, as well as a single external edge from the original vertex. Now, we consider all possible Eulerian circuits through the subgraph resulting from the 1-4 move. To reduce the number of options to be considered, the following result is helpful.

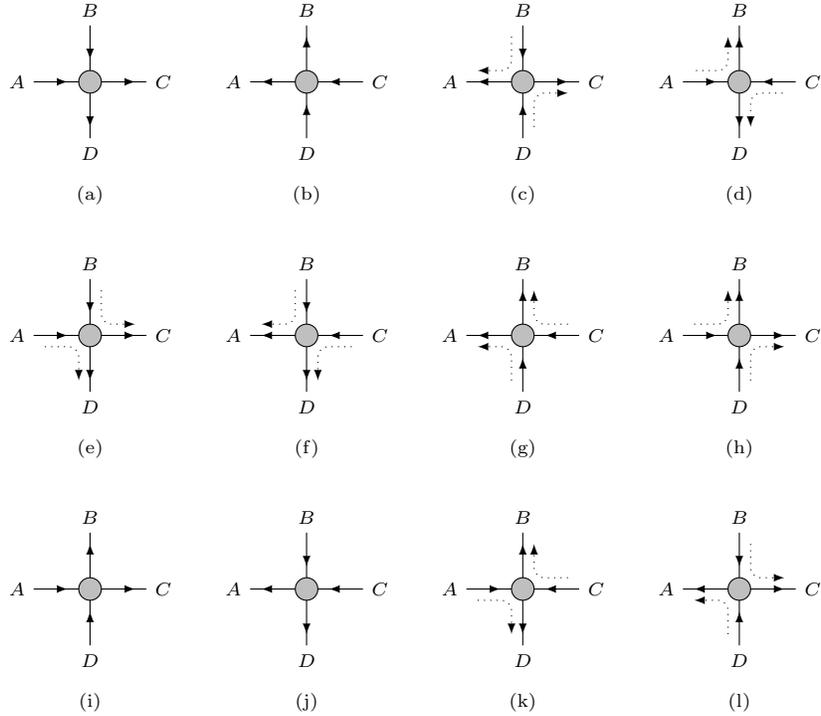
\begin{figure}[hbt]
\centering
\begin{tikzpicture}[> = latex, font = \scriptsize]
\matrix[column sep = 0.5 cm, row sep = 0.5 cm]{


\draw [fill = gray!50] (0, 0) circle (0.15);


\begin{scope}[decoration = {markings, mark = at position 0.75 with {\arrow{latex}}}]

	\draw [postaction = {decorate}] (0.15, 0) -- (0.75, 0) node [right] {$C$};
	\draw [postaction = {decorate}] (-0.75, 0) node [left] {$A$} -- (-0.15, 0);
	\draw [postaction = {decorate}] (0, -0.15) -- (0, -0.75) node [below] {$D$};
	\draw [postaction = {decorate}] (0, 0.75) node [above] {$B$} -- (0, 0.15);

\end{scope}


\node at (0, -1.5) {(a)};

&


\draw [fill = gray!50] (0, 0) circle (0.15);


\begin{scope}[decoration = {markings, mark = at position 0.75 with {\arrow{latex}}}]

	\draw [postaction = {decorate}] (0.75, 0) node [right] {$C$} -- (0.15, 0);
	\draw [postaction = {decorate}] (-0.15, 0) -- (-0.75, 0) node [left] {$A$};
	\draw [postaction = {decorate}] (0, -0.75) node [below] {$D$} -- (0, -0.15);
	\draw [postaction = {decorate}] (0, 0.15) -- (0, 0.75) node [above] {$B$};

\end{scope}


\node at (0, -1.5) {(b)};

&


\draw [fill = gray!50] (0, 0) circle (0.15);


\begin{scope}[decoration = {markings, mark = at position 0.75 with {\arrow{latex}}}]

	\draw [postaction = {decorate}] (0.15, 0) -- (0.75, 0) node [right] {$C$};
	\draw [postaction = {decorate}] (-0.15, 0) -- (-0.75, 0) node [left] {$A$};
	\draw [postaction = {decorate}] (0, -0.75) node [below] {$D$} -- (0, -0.15);
	\draw [postaction = {decorate}] (0, 0.75) node [above] {$B$} -- (0, 0.15);

\end{scope}


\begin{scope}[->, dotted, rounded corners]

	\draw (-0.15, 0.6) -- (-0.15, 0.15) -- (-0.6, 0.15);
	\draw (0.15, -0.6) -- (0.15, -0.15) -- (0.6, -0.15);

\end{scope}


\node at (0, -1.5) {(c)};

&


\draw [fill = gray!50] (0, 0) circle (0.15);


\begin{scope}[decoration = {markings, mark = at position 0.75 with {\arrow{latex}}}]

	\draw [postaction = {decorate}] (0.75, 0) node [right] {$C$} -- (0.15, 0);
	\draw [postaction = {decorate}] (-0.75, 0) node [left] {$A$} -- (-0.15, 0);
	\draw [postaction = {decorate}] (0, -0.15) -- (0, -0.75) node [below] {$D$};
	\draw [postaction = {decorate}] (0, 0.15) -- (0, 0.75) node [above] {$B$};

\end{scope}


\begin{scope}[<-, dotted, rounded corners]

	\draw (-0.15, 0.6) -- (-0.15, 0.15) -- (-0.6, 0.15);
	\draw (0.15, -0.6) -- (0.15, -0.15) -- (0.6, -0.15);

\end{scope}


\node at (0, -1.5) {(d)};

\\


\draw [fill = gray!50] (0, 0) circle (0.15);


\begin{scope}[decoration = {markings, mark = at position 0.75 with {\arrow{latex}}}]

	\draw [postaction = {decorate}] (0.15, 0) -- (0.75, 0) node [right] {$C$};
	\draw [postaction = {decorate}] (-0.75, 0) node [left] {$A$} -- (-0.15, 0);
	\draw [postaction = {decorate}] (0, -0.15) -- (0, -0.75) node [below] {$D$};
	\draw [postaction = {decorate}] (0, 0.75) node [above] {$B$} -- (0, 0.15);

\end{scope}


\begin{scope}[->, dotted, rounded corners]

	\draw (0.15, 0.6) -- (0.15, 0.15) -- (0.6, 0.15);
	\draw (-0.6, -0.15) -- (-0.15, -0.15) -- (-0.15, -0.6);

\end{scope}


\node at (0, -1.5) {(e)};

&


\draw [fill = gray!50] (0, 0) circle (0.15);


\begin{scope}[decoration = {markings, mark = at position 0.75 with {\arrow{latex}}}]

	\draw [postaction = {decorate}] (0.75, 0) node [right] {$C$} -- (0.15, 0);
	\draw [postaction = {decorate}] (-0.15, 0) -- (-0.75, 0) node [left] {$A$};
	\draw [postaction = {decorate}] (0, -0.15) -- (0, -0.75) node [below] {$D$};
	\draw [postaction = {decorate}] (0, 0.75) node [above] {$B$} -- (0, 0.15);

\end{scope}


\begin{scope}[->, dotted, rounded corners]

	\draw (-0.15, 0.6) -- (-0.15, 0.15) -- (-0.6, 0.15);
	\draw (0.6, -0.15) -- (0.15, -0.15) -- (0.15, -0.6);

\end{scope}


\node at (0, -1.5) {(f)};

&


\draw [fill = gray!50] (0, 0) circle (0.15);


\begin{scope}[decoration = {markings, mark = at position 0.75 with {\arrow{latex}}}]

	\draw [postaction = {decorate}] (0.75, 0) node [right] {$C$} -- (0.15, 0);
	\draw [postaction = {decorate}] (-0.15, 0) -- (-0.75, 0) node [left] {$A$};
	\draw [postaction = {decorate}] (0, -0.75) node [below] {$D$} -- (0, -0.15);
	\draw [postaction = {decorate}] (0, 0.15) -- (0, 0.75) node [above] {$B$};

\end{scope}


\begin{scope}[->, dotted, rounded corners]

	\draw (-0.15, -0.6) -- (-0.15, -0.15) -- (-0.6, -0.15);
	\draw (0.6, 0.15) -- (0.15, 0.15) -- (0.15, 0.6);

\end{scope}


\node at (0, -1.5) {(g)};

&


\draw [fill = gray!50] (0, 0) circle (0.15);


\begin{scope}[decoration = {markings, mark = at position 0.75 with {\arrow{latex}}}]

	\draw [postaction = {decorate}] (0.15, 0) -- (0.75, 0) node [right] {$C$};
	\draw [postaction = {decorate}] (-0.75, 0) node [left] {$A$} -- (-0.15, 0);
	\draw [postaction = {decorate}] (0, -0.75) node [below] {$D$} -- (0, -0.15);
	\draw [postaction = {decorate}] (0, 0.15) -- (0, 0.75) node [above] {$B$};

\end{scope}


\begin{scope}[->, dotted, rounded corners]

	\draw (0.15, -0.6) -- (0.15, -0.15) -- (0.6, -0.15);
	\draw (-0.6, 0.15) -- (-0.15, 0.15) -- (-0.15, 0.6);

\end{scope}


\node at (0, -1.5) {(h)};

\\	


\draw [fill = gray!50] (0, 0) circle (0.15);


\begin{scope}[decoration = {markings, mark = at position 0.75 with {\arrow{latex}}}]

	\draw [postaction = {decorate}] (0.15, 0) -- (0.75, 0) node [right] {$C$};
	\draw [postaction = {decorate}] (-0.75, 0) node [left] {$A$} -- (-0.15, 0);
	\draw [postaction = {decorate}] (0, -0.75) node [below] {$D$} -- (0, -0.15);
	\draw [postaction = {decorate}] (0, 0.15) -- (0, 0.75) node [above] {$B$};

\end{scope}


\node at (0, -1.5) {(i)};

&


\draw [fill = gray!50] (0, 0) circle (0.15);


\begin{scope}[decoration = {markings, mark = at position 0.75 with {\arrow{latex}}}]

	\draw [postaction = {decorate}] (0.75, 0) node [right] {$C$} -- (0.15, 0);
	\draw [postaction = {decorate}] (-0.15, 0) -- (-0.75, 0) node [left] {$A$};
	\draw [postaction = {decorate}] (0, -0.15) -- (0, -0.75) node [below] {$D$};
	\draw [postaction = {decorate}] (0, 0.75) node [above] {$B$} -- (0, 0.15);

\end{scope}


\node at (0, -1.5) {(j)};

&


\draw [fill = gray!50] (0, 0) circle (0.15);


\begin{scope}[decoration = {markings, mark = at position 0.75 with {\arrow{latex}}}]

	\draw [postaction = {decorate}] (0.75, 0) node [right] {$C$} -- (0.15, 0);
	\draw [postaction = {decorate}] (-0.75, 0) node [left] {$A$} -- (-0.15, 0);
	\draw [postaction = {decorate}] (0, -0.15) -- (0, -0.75) node [below] {$D$};
	\draw [postaction = {decorate}] (0, 0.15) -- (0, 0.75) node [above] {$B$};

\end{scope}


\begin{scope}[->, dotted, rounded corners]

	\draw (-0.6, -0.15) -- (-0.15, -0.15) -- (-0.15, -0.6);
	\draw (0.6, 0.15) -- (0.15, 0.15) -- (0.15, 0.6);

\end{scope}


\node at (0, -1.5) {(k)};

&


\draw [fill = gray!50] (0, 0) circle (0.15);


\begin{scope}[decoration = {markings, mark = at position 0.75 with {\arrow{latex}}}]

	\draw [postaction = {decorate}] (0.15, 0) -- (0.75, 0) node [right] {$C$};
	\draw [postaction = {decorate}] (-0.15, 0) -- (-0.75, 0) node [left] {$A$};
	\draw [postaction = {decorate}] (0, -0.75) node [below] {$D$} -- (0, -0.15);
	\draw [postaction = {decorate}] (0, 0.75) node [above] {$B$} -- (0, 0.15);

\end{scope}


\begin{scope}[->, dotted, rounded corners]

	\draw (-0.15, -0.6) -- (-0.15, -0.15) -- (-0.6, -0.15);
	\draw (0.15, 0.6) -- (0.15, 0.15) -- (0.6, 0.15);

\end{scope}


\node at (0, -1.5) {(l)};

\\
};
\end{tikzpicture}
\caption{\label{lemma}Each of the four situations in a given row will appear the same number of times in the collection of Eulerian circuits for a given graph diagram. The vertex state is irrelevant for this result.}
\end{figure}

\begin{proposition}
\label{4-similar}
	For a given vertex in the graph diagram of a knotted 4-regular pseudograph, suppose there are $N$ Eulerian circuits where one of the four situations in one of the three rows in Figure \ref{lemma} appears. Then there will be $N$ circuits where each of the other three situations occur in the same row.
\end{proposition}
{\it Proof:} For a given edge, there are three choices of the other edge it connects to in the Eulerian circuit. For each of these choices, the other two edges must be automatically connected as well. Finally, with all three choices, each pair of edges has two choices for which way the Eulerian circuit passes through the edges. Thus, in total, there are 12 ways for the circuit to pass through the vertex.

Suppose the vertex is labeled as $V$. We consider the cases (a) -- (d) in Figure \ref{lemma} first. For those four situations, the Eulerian circuit passes through the vertex $V$ so that, through the remainder of the graph diagram,  $A$ is connected to $D$, and $B$ is connected to $C$ (with the order depending on the orientations of the edges). In other words, following the orientations of the edges in the graph diagram, there is a path from $A$ to $D$ (or vice versa), and similiarly for $B$ and $C$. Thus, starting with case (a) and the edge orientations given by its Eulerian circuit, the other three situations are obtained by the following means.

\begin{enumerate}

	\item Case (b) is obtained from case (a) by reversing all edge orientations on the graph diagram.

	\item Case (c) is obtained from case (a) by reversing the edge orientations on the loop $AVDA$. Since the edges $A$ and $D$ are connected by the rest of the graph diagram without passing through the vertex $V$ first, this does not reverse all edges on the graph diagram.

	\item Case (d) is obtained from case (a) by reversing the edge orientations on the loop $BVCB$. Again, $B$ and $C$ are connected, so this only reverses some of the edges of the graph diagram, not all.

\end{enumerate}
The proof for the other two rows in Figure \ref{lemma} follows a similar method, where situations (e) -- (h) occur when $A, C$ and $B, D$ are the connected pairs, and (i) -- (l) occur when $A, B$ and $C, D$ are the connected pairs. $\blacksquare$

In addition, Proposition \ref{4-similar} implies the following result.

\begin{corollary}
	The total number of Eulerian circuits for a knotted 4-regular pseudograph is a multiple of four.
\end{corollary}

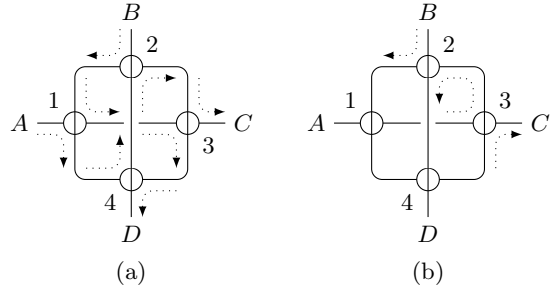
\begin{figure}[hbt]
\centering
\begin{tikzpicture}[> = latex]
\matrix[column sep = 0.5 cm]{


	\draw (-1.25, 0) node [left] {$A$} -- (1.25, 0) node [right] {$C$};

	\draw [fill = white] (-0.75, 0) circle (0.15) node [above left = 0.25 em] {1};
	\draw [fill = white] (0, 0.75) circle (0.15) node [above right = 0.25 em] {2};
	\draw [fill = white] (0.75, 0) circle (0.15) node [below right = 0.25 em] {3};
	\draw [fill = white] (0, -0.75) circle (0.15) node [below left = 0.25 em] {4};

	\draw [rounded corners] (-0.15, -0.75) -- (-0.75, -0.75) -- (-0.75, 0.75) -- (-0.15, 0.75)
		(0.15, -0.75) -- (0.75, -0.75) -- (0.75, 0.75) -- (0.15, 0.75);

	\draw [draw = white, double = black, double distance between line centers = 3 pt, line width = 2.6 pt] (0, -0.6) -- (0, 0.6);

	\draw (0, 1.25) node [above] {$B$} -- (0, 0.6);
	\draw (0, -0.6) -- (0, -1.25) node [below] {$D$};


	\begin{scope}[->, dotted, rounded corners]
	
	
		\draw (-0.6, 0.6) -- (-0.6, 0.15) -- (-0.15, 0.15);
		\draw (-1.25, -0.15) -- (-0.9, -0.15) -- (-0.9, -0.6);
		
		
		\draw (-0.15, 1.25) -- (-0.15, 0.9) -- (-0.6, 0.9);
		\draw (0.15, 0.15) -- (0.15, 0.6) -- (0.6, 0.6);
		
		
		\draw (0.9, 0.6) -- (0.9, 0.15) -- (1.25, 0.15);
		\draw (0.15, -0.15) -- (0.6, -0.15) -- (0.6, -0.6);
		
		
		\draw (-0.6, -0.6) -- (-0.15, -0.6) -- (-0.15, -0.15);
		\draw (0.6, -0.9) -- (0.15, -0.9) -- (0.15, -1.25);

	\end{scope}


	\node at (0, -2) {(a)};

&


	\draw (-1.25, 0) node [left] {$A$} -- (1.25, 0) node [right] {$C$};

	\draw [fill = white] (-0.75, 0) circle (0.15) node [above left = 0.25 em] {1};
	\draw [fill = white] (0, 0.75) circle (0.15) node [above right = 0.25 em] {2};
	\draw [fill = white] (0.75, 0) circle (0.15) node [above right = 0.25 em] {3};
	\draw [fill = white] (0, -0.75) circle (0.15) node [below left = 0.25 em] {4};

	\draw [rounded corners] (-0.15, -0.75) -- (-0.75, -0.75) -- (-0.75, 0.75) -- (-0.15, 0.75)
		(0.15, -0.75) -- (0.75, -0.75) -- (0.75, 0.75) -- (0.15, 0.75);

	\draw [draw = white, double = black, double distance between line centers = 3 pt, line width = 2.6 pt] (0, -0.6) -- (0, 0.6);

	\draw (0, 1.25) node [above] {$B$} -- (0, 0.6);
	\draw (0, -0.6) -- (0, -1.25) node [below] {$D$};


	\begin{scope}[->, dotted, rounded corners]
		
		
		\draw (-0.15, 1.25) -- (-0.15, 0.9) -- (-0.6, 0.9);
		
		
		\draw (0.9, -0.6) -- (0.9, -0.15) -- (1.25, -0.15);

		
		\draw (0.25, 0.15) -- (0.6, 0.15) -- (0.6, 0.6) -- (0.15, 0.6) -- (0.15, 0.25);

	\end{scope}


	\node at (0, -2) {(b)};

\\
};
\end{tikzpicture}
\caption{\label{4-vert-sym}Each case shown here is obtained from the other under the exchange $3 \leftrightarrow 4, C \leftrightarrow D$. In case (a), the Eulerian circuit passes through the vertex as $A \to C, B \to D$, while for case (b), the circuit is $A \to D, B \to C$. Note that in case (b), both vertices 1 and 4 are now proper vertices. }
\end{figure}

The upshot of this discussion is that there are only two cases to consider for the starting vertex of the 1-4 move, and the orientations of the incident edges. Proposition \ref{4-similar} means we only need to consider one case out of each row in Figure \ref{lemma}. Additionally, the third row is obtained from the first by a vertical reflection of the vertex and its edges. In fact, since the resulting subgraphs of four vertices after the 1-4 move are the same as abstract graphs, then there is a direct mapping of edge orientations between the two cases. An example of this mapping is shown in Figure \ref{4-vert-sym}. In case (a) of Figure \ref{4-vert-sym}, the Eulerian circuit passes through the subgraph as $A \to 1 \to 4 \to 2 \to 3 \to C,$ and $B \to 1 \to 3 \to 4 \to D$ This is a result of a 1-4 move on the vertex (a) of Figure \ref{lemma}. On the other hand, acting on vertex (e) of Figure \ref{lemma} by a 1-4 move gives case (b) of Figure \ref{4-vert-sym}. For this case, one exchanges 3 with 4, and $C$ with $D$, in the edge orderings given for case (a), so that the Eulerian circuit now passes through the subgraph as $A \to 1 \to 3 \to 2 \to 4 \to D$, and $B \to 1 \to 4 \to 3 \to C$. This provides a mapping between a circuit passing through the subgraph between $A \to C, B \to D$ for case (a), and $A \to D, B \to C$ for case (b). It is important to note that these do not give a matching family of Eulerian circuits, in the sense of Proposition \ref{4-similar}, but instead are mappings between two different ways of connecting the circuit through the subgraph.

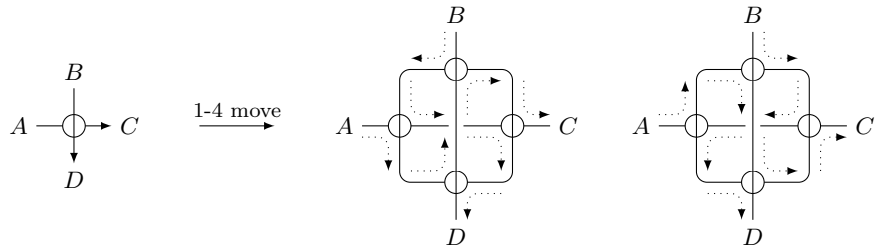
\begin{figure}[hbt]
\centering
\begin{tikzpicture}[> = latex]
\matrix[column sep = 0.5 cm]{


	\draw [->] (-0.5, 0) node [left] {$A$} -- (0.5, 0) node [right] {$C$};
	\draw [fill = white] (0, 0) circle (0.15);
	\draw [<-] (0, -0.5) node [below] {$D$} -- (0, 0.5) node [above] {$B$};

&

	\draw [->, font = \footnotesize] (0, 0) -- node [above] {1-4 move} (1, 0);

&


	\draw (-1.25, 0) node [left] {$A$} -- (1.25, 0) node [right] {$C$};

	\draw [fill = white] (-0.75, 0) circle (0.15);
	\draw [fill = white] (0, 0.75) circle (0.15);
	\draw [fill = white] (0.75, 0) circle (0.15);
	\draw [fill = white] (0, -0.75) circle (0.15);

	\draw [rounded corners] (-0.15, -0.75) -- (-0.75, -0.75) -- (-0.75, 0.75) -- (-0.15, 0.75)
		(0.15, -0.75) -- (0.75, -0.75) -- (0.75, 0.75) -- (0.15, 0.75);

	\draw [draw = white, double = black, double distance between line centers = 3 pt, line width = 2.6 pt] (0, -0.6) -- (0, 0.6);

	\draw (0, 1.25) node [above] {$B$} -- (0, 0.6);
	\draw (0, -0.6) -- (0, -1.25) node [below] {$D$};

	
	\begin{scope}[->, dotted, rounded corners]
	
		\draw (-0.6, 0.6) -- (-0.6, 0.15) -- (-0.15, 0.15);
		\draw (-1.25, -0.15) -- (-0.9, -0.15) -- (-0.9, -0.6);
		
		\draw (-0.15, 1.25) -- (-0.15, 0.9) -- (-0.6, 0.9);
		\draw (0.15, 0.15) -- (0.15, 0.6) -- (0.6, 0.6);
		
		\draw (0.9, 0.6) -- (0.9, 0.15) -- (1.25, 0.15);
		\draw (0.15, -0.15) -- (0.6, -0.15) -- (0.6, -0.6);
		
		\draw (-0.6, -0.6) -- (-0.15, -0.6) -- (-0.15, -0.15);
		\draw (0.6, -0.9) -- (0.15, -0.9) -- (0.15, -1.25);
	
	\end{scope}

&


	\draw (-1.25, 0) node [left] {$A$} -- (1.25, 0) node [right] {$C$};

	\draw [fill = white] (-0.75, 0) circle (0.15);
	\draw [fill = white] (0, 0.75) circle (0.15);
	\draw [fill = white] (0.75, 0) circle (0.15);
	\draw [fill = white] (0, -0.75) circle (0.15);

	\draw [rounded corners] (-0.15, -0.75) -- (-0.75, -0.75) -- (-0.75, 0.75) -- (-0.15, 0.75)
		(0.15, -0.75) -- (0.75, -0.75) -- (0.75, 0.75) -- (0.15, 0.75);

	\draw [draw = white, double = black, double distance between line centers = 3 pt, line width = 2.6 pt] (0, -0.6) -- (0, 0.6);

	\draw (0, 1.25) node [above] {$B$} -- (0, 0.6);
	\draw (0, -0.6) -- (0, -1.25) node [below] {$D$};

	
	\begin{scope}[->, dotted, rounded corners]
	
		\draw (-1.25, 0.15) -- (-0.9, 0.15) -- (-0.9, 0.6);
		\draw (-0.15, -0.15) -- (-0.6, -0.15) -- (-0.6, -0.6);
	
		\draw (0.15, 1.25) -- (0.15, 0.9) -- (0.6, 0.9);
		\draw (-0.6, 0.6) -- (-0.15, 0.6) -- (-0.15, 0.15);
		
		\draw (0.6, 0.6) -- (0.6, 0.15) -- (0.15, 0.15);
		\draw (0.9, -0.6) -- (0.9, -0.15) -- (1.25, -0.15);
		
		\draw (0.15, -0.15) -- (0.15, -0.6) -- (0.6, -0.6);
		\draw (-0.6, -0.9) -- (-0.15, -0.9) -- (-0.15, -1.25);
	
	\end{scope}

\\
};
\end{tikzpicture}
\caption{\label{4-vert-switch}The two cases where the crossing type of the induced knot diagram changes after a 1-4 move. There are 20 other possible ways for the Eulerian circuit to pass through the resulting subgraph, which have the same crossing type as the original vertex on the left side.}
\end{figure}

Thus, any circuit passing through the subgraph obtained after the 1-4 move can be obtained by an exchange just discussed. We shall then only use the starting situation given as the left-hand side of Figure \ref{4-vert-switch}; all other cases are given by the appropriate mappings. Note that the induced knot diagram of the subgraph for the starting vertex in Figure \ref{4-vert-switch} has the edge $B \to D$ passing over the edge $A \to C$. After the 1-4 move, there are 22 possible ways for the Eulerian circuit to pass through the subgraph from $A \to C, B \to D$. In 20 of these circuits, the resulting induced knot diagram is equivalent (up to Reidemeister moves) to the same passage of the edge $B \to D$ over that of $A \to C$. However, for the remaining two cases, shown to the right of Figure \ref{4-vert-switch}, the crossing type is reversed. In other words, it is now the edge $A \to C$ that passes over $B \to D$. By the argument given previously, with the mapping exemplified by Figure \ref{4-vert-sym}, there are also two cases for the situation with $A \to D, B \to C$ where the induced knot diagram changes. In these cases, before the 1-4 move, there is no crossing at all. Afterwards, however, there are two crossings after the Pachner move. One of these situations is shown as case (b) in Figure \ref{4-vert-sym} -- the part of the circuit $A \to D$ first passes under the portion $B \to C$ at vertex 1, and then over it at vertex 4. In either alternative in Figure \ref{4-vert-sym} and related Eulerian circuits, the 1-4 move gives a {\it 2-move} on the induced knot diagram -- either switching the crossing type, or else adding two opposite crossings to two edges originally disjoint. In general, this induced knot diagram is equivalent to a different knot than was present before the Pachner move.

\begin{figure}[hbt]
\centering
\begin{tikzpicture}[> = latex]
\matrix[column sep = 0.5 cm]{


	\draw [rounded corners] (-0.15, 0) -- (-1, 0) -- (-1, -0.5) -- (0.5, -0.5) -- (0.5, -0.15);


	\draw [fill = white] (-0.5, 0) circle (0.15);


	\draw [rounded corners] (0.5, 0.15) -- (0.5, 0.3) -- (0, 0.3) -- (0, -0.3) -- (-0.5, -0.3) -- (-0.5, 0.5) -- (1, 0.5) -- (1, 0) -- (0.15, 0);


	\node at (0, -1.75) {$3^\ell 5^+ 1^+$};

&

	\draw [->, font = \footnotesize] (0, 0) -- node [above] {1-4 move} (1, 0);

&


	\draw [rounded corners] (1.5, -0.15) -- (1.5, -1.25) -- (-1, -1.25) -- (-1, 0) -- (0.85, 0);

	\draw [fill = white] (-0.5, 0) circle (0.15);
	\draw [fill = white] (0, 0.5) circle (0.15);
	\draw [fill = white] (0.5, 0) circle (0.15);
	\draw [fill = white] (0, -0.5) circle (0.15);

	\draw [rounded corners] (-0.15, -0.5) -- (-0.5, -0.5) -- (-0.5, 0.5) -- (-0.15, 0.5)
		(0.15, -0.5) -- (0.5, -0.5) -- (0.5, 0.5) -- (0.15, 0.5);

	\draw [draw = white, double = black, double distance between line centers = 3 pt, line width = 2.6 pt] (0, -0.35) -- (0, 0.35);

	\draw [rounded corners] (1.15, 0) -- (2, 0) -- (2, 1) -- (0, 1) -- (0, 0.35);
	\draw [rounded corners] (0, -0.35) -- (0, -1) -- (1, -1) -- (1, 0.5) -- (1.5, 0.5) -- (1.5, 0.15);


	\node at (0.5, -1.75) {$5^\ell 7^- 13^\ell 11^\ell 3^u 15^+ 1^+ 9^+$};

&


	\draw [gray, very thick, rounded corners] (1.15, 0) -- (2, 0) -- (2, 1) -- (0, 1) -- (0, 0.65)
		(0, -0.65) -- (0, -1) -- (1, -1) -- (1, 0.5) -- (1.5, 0.5) -- (1.5, 0.15)
		(1.5, -0.15) -- (1.5, -1.25) -- (-1, -1.25) -- (-1, 0) -- (-0.65, 0)
		(-0.5, -0.15) -- (-0.5, -0.5) -- (-0.15, -0.5)
		(0, -0.35) -- (0, 0.35)
		(0.15, 0.5) -- (0.5, 0.5) -- (0.5, 0.15)
		(0.65, 0) -- (0.85, 0)
		(0.15, -0.5) -- (0.5, -0.5) -- (0.5, -0.15)
		(0.35, 0) -- (0.15, 0)
		(-0.15, 0) -- (-0.35, 0)
		(-0.5, 0.15) -- (-0.5, 0.5) -- (-0.15, 0.5);
	\draw [gray, very thick] (-0.65, 0) arc (180 : 270 : 0.15)
		(-0.15, -0.5) arc (180 : 90 : 0.15)
		(0, 0.35) arc (270 : 360 : 0.15)
		(0.5, 0.15) arc (90 : 0 : 0.15)
		(0, -0.65) arc (270 : 360 : 0.15)
		(0.5, -0.15) arc (270 : 180 : 0.15)
		(-0.35, 0) arc (0 : 90 : 0.15)
		(-0.15, 0.5) arc (180 : 90 : 0.15);

\\
};
\end{tikzpicture}
\caption{\label{1-4-ex}The graphs $3^\ell 5^+ 1^+$ and $5^\ell 7^- 13^\ell 11^\ell 3^u 15^+ 1^+ 9^+$. The latter graph is shown directly after the 1-4 move, not in the form implied by its DT sequence. To the right is a possible Eulerian circuit through the resulting graph, whose induced knot diagram corresponds to a trefoil knot.}
\end{figure}
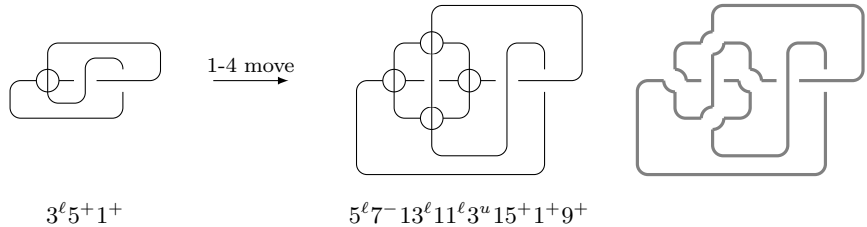

As an example of this discussion, we consider a 1-4 move acting on the graph $3^\ell 5^+ 1^+$. This is shown in Figure \ref{1-4-ex}. To start with, there are four Eulerian circuits through the graph. This can be easily seen by using Proposition \ref{4-similar}, starting from the proper form of the graph and using (a) -- (d) in Figure \ref{lemma}. Each of these circuits gives an induced knot diagram equivalent to the unknot. Thus, choosing the same Takasaki kei $\mathbb{Z}_3$ used in Section \ref{quandle-poly} as our quandle (with operations given in Table \ref{kei}), then
\[
	\Psi_{\mathbb{Z}_3} (3^\ell 5^+ 1^+) = 4u^3
\]
Under the 1-4 move, the graph on the right-hand side of Figure \ref{1-4-ex} is obtained. This graph diagram has 88 total Eulerian circuits, 22 for each of the possibilities (a) -- (d) shown in Figure \ref{lemma} of the original vertex. For each configuration of the original vertex, the final 4-vertex subgraph has 20 Eulerian circuits that preserve the original crossing type. On the other hand, the final two circuits for each vertex possibility change the induced knot diagram to give the trefoil knot. The induced knot diagram for one of these circuits is shown to the right in Figure \ref{1-4-ex}. As given in Section \ref{quandle-poly}, the trefoil has a total of nine quandle colorings for the kei $\mathbb{Z}_3$, so that the quandle counting polynomial for the final graph is given by
\[
	\Psi_{\mathbb{Z}_3} (5^\ell 7^- 13^\ell 11^\ell 3^u 15^+ 1^+ 9^+) = 80u^3 + 8u^9
\]
The 1-4 move has increased the number of quandle colorings possible in all induced knot diagrams, so the polynomial $\Psi_{\mathbb{Z}_3}$ has more terms than before. Mathematical relations such as these may help quantify the partial order on the space of knotted graphs under the Pachner moves.

At this point, we return to a comment made at the end of Section \ref{notation}, where some issues with finding the minimal DT sequence for a graph where discussed. The situation above gives an example of this -- namely, how to describe the graph resulting after a 1-4 move on the graph $3^\ell 5^+ 1^+$. In Figure \ref{1-4-ex}, the graph after a 1-4 move is shown so that the insertion of the 4-vertex subgraph is obvious. However, in this form, the graph diagram is not in proper form, and thus not amenable to finding the DT sequence. To do so, it is necessary to use the process described to prove Proposition \ref{DT-seq-prop}. However, it is not obvious which Eulerian circuit to base these moves on. For the graph considered here, since there are no other vertices in the graph, the choice was made to use the circuit which has the smallest number of improper vertices. This chosen Eulerian circuit has a single improper vertex, of the form shown in Figure \ref{bad-vert-2}. Thus a RV move was necessary to change the vertex to a proper one, giving one additional crossing in the graph diagram. After this vertex rotation, the DT sequence coming lowest in lexicographical order was obtained. It is not obvious that this process is satisfactory in general, especially when there are other vertices in the graph unaffected by the 1-4 move. This question is left for future work, when a larger census of knotted 4-regular graphs has been obtained.

\subsection{The 2-3 move}
\label{2-3-sect}

\begin{figure}[hbt]
\centering
\begin{tikzpicture}[> = latex]
\matrix[column sep = 0.5 cm, row sep = 0.5 cm]{


	\draw (-1, 0) node [left] {$B$} -- (1, 0) node [right] {$B'$};

	\draw [fill = white] (-0.5, 0) circle (0.15);
	\draw [fill = white] (0.5, 0) circle (0.15);

	\draw (-0.5, 0.5) node [above] {$A$} -- (-0.5, -0.5) node [below] {$C$};
	\draw (0.5, 0.5) node [above] {$A'$} -- (0.5, -0.5) node [below] {$C'$};

&

	\begin{scope}[->, font = \footnotesize]

		\draw (0, 0.25) -- node [above] {2-3 move} (1, 0.25);
		\draw (1, -0.25) -- node [below] {3-2 move} (0, -0.25);

	\end{scope}

&


	\draw (-0.5, 0) node [left] {$B$} -- (1.25, 0) node [right] {$B'$};
	\draw [fill = white] (0.75, 0) circle (0.15);

	\draw [rounded corners] (-0.5, 0.5) node [left] {$A$} -- (0.75, 0.5) -- (0.75, -0.5) -- (-0.5, -0.5) node [left] {$C$};

	\draw [fill = white] (0, 0.5) circle (0.15);
	\draw [fill = white] (0, -0.5) circle (0.15);

	\draw (0, 1) node [above] {$A'$} -- (0, 0.35);
	\draw [draw = white, double = black, double distance between line centers = 3 pt, line width = 2.6 pt] (0, 0.35)  -- (0, -0.35);
	\draw (0, -1) node [below] {$C'$} -- (0, -0.35);

\\


	\draw (-0.5, 0.5) node [above] {$A$} -- (-0.5, -0.5) node [below] {$C$};
	\draw (0.5, 0.5) node [above] {$A'$} -- (0.5, -0.5) node [below] {$C'$};

	\draw [fill = white] (-0.5, 0) circle (0.15);
	\draw [fill = white] (0.5, 0) circle (0.15);

	\draw (-1, 0) node [left] {$B$} -- (1, 0) node [right] {$B'$};

&

	\begin{scope}[->, font = \footnotesize]

		\draw (0, 0.25) -- node [above] {2-3 move} (1, 0.25);
		\draw (1, -0.25) -- node [below] {3-2 move} (0, -0.25);

	\end{scope}

&


	\draw (0, 1) node [above] {$A'$} -- (0, -1) node [below] {$C'$};

	\draw [fill = white] (0, 0.5) circle (0.15);
	\draw [fill = white] (0.75, 0) circle (0.15);
	\draw [fill = white] (0, -0.5) circle (0.15);

	\draw [draw = white, double = black, double distance between line centers = 3 pt, line width = 2.6 pt] (-0.5, 0) node [left] {$B$} -- (0.6, 0);
	\draw [rounded corners] (-0.5, 0.5) node [left] {$A$} -- (0.75, 0.5) -- (0.75, 0.15)
		(0.75, -0.15) -- (0.75, -0.5) -- (-0.5, -0.5) node [left] {$C$};
	\draw (0.6, 0) -- (1.25, 0) node [right] {$B'$};

\\
};
\end{tikzpicture}
\caption{\label{2-3}2-3 and 3-2 Pachner graph moves.}
\end{figure}
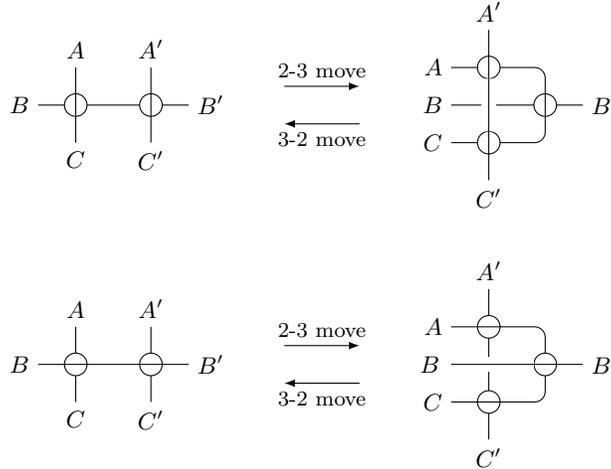

Now we turn to the Pachner 2-3 move, which acts on two vertices sharing an edge, as shown in Figure \ref{2-3}. The result of the move is a subgraph of three vertices, all connected together, along with each of the new vertices also having two of the external edges incident to it. The 3-2 move acts as the inverse, removing the 3-cycle of vertices and replacing it with a single edge. As discussed previously in Section \ref{evo-moves}, it is assumed that there are no external edges passing through the 3-cycle that would block the 3-2 move. In other words, the linking number of any external edge with the three edges joining the vertices must be zero, so the external edge can be moved outside the subgraph where the 3-2 move takes place.

Next, we consider all the possible ways that the Eulerian circuit can pass through the 2-vertex subgraph where the 2-3 Pachner move will act. Starting from the 2-vertex system, the external edges are divided into two sets, $\{A, B, C\}$ for those on the left, and $\{A', B', C'\}$ on the right. There are five choices for the edge to connect to the first edge, three for the second, and one choice for the third; this ignores the addition of an orientation for each of these connections. Thus, there are 15 different ways for the Eulerian circuit to connect these edges. However, six out of these 15 choices would result in the circuit passing through the edge linking the two vertices more than once, so there are only nine possible ways to connect the external edges in the Eulerian circuit for the 2-vertex subgraph. Each of these possibilities has $2^3 = 8$ ways for the circuit to pass through the subgraph: two choices for the orientation of each of the connections. Together, this means there are 72 possible ways for an Eulerian circuit to pass through the 2-vertex subgraph.

\begin{figure}[hbt]
\centering
\begin{tikzpicture}[> = latex]
\matrix[column sep = 0.5 cm, row sep = 0.5 cm]{


	\draw [->] (-1.25, 0) node [left] {$B$} -- (1.25, 0) node [right] {$B'$};

	\draw [fill = white] (-0.5, 0) circle (0.15);
	\draw [fill = white] (0.5, 0) circle (0.15);

	\draw [->] (-0.5, 0.75) node [above] {$A$} -- (-0.5, -0.75) node [below] {$C$};
	\draw [->] (0.5, 0.75) node [above] {$A'$} -- (0.5, -0.75) node [below] {$C'$};

&

	\draw [->, font = \footnotesize] (0, 0) -- node [above] {2-3 move} (1, 0);

&


	\draw [->] (-0.5, 0) node [left] {$B$} -- (1.25, 0) node [right] {$B'$};
	\draw [fill = white] (0.75, 0) circle (0.15);

	\draw [rounded corners] (-0.5, 0.75) node [left] {$A$} -- (0.75, 0.75) -- (0.75, -0.75) -- (-0.5, -0.75) node [left] {$C$};

	\draw [fill = white] (0, 0.75) circle (0.15);
	\draw [fill = white] (0, -0.75) circle (0.15);

	\draw (0, 1.25) node [above] {$A'$} -- (0, 0.35);
	\draw [draw = white, double = black, double distance between line centers = 3 pt, line width = 2.6 pt] (0, 0.35)  -- (0, -0.35);
	\draw [<-] (0, -1.25) node [below] {$C'$} -- (0, -0.35);

&


	\draw [->] (-0.5, 0) node [left] {$B$} -- (1.25, 0) node [right] {$B'$};
	\draw [fill = white] (0.75, 0) circle (0.15);

	\draw [rounded corners] (-0.5, 0.75) node [left] {$A$} -- (0.75, 0.75) -- (0.75, -0.75) -- (-0.5, -0.75) node [left] {$C$};

	\draw [fill = white] (0, 0.75) circle (0.15);
	\draw [fill = white] (0, -0.75) circle (0.15);

	\draw (0, 1.25) node [above] {$A'$} -- (0, 0.35);
	\draw [draw = white, double = black, double distance between line centers = 3 pt, line width = 2.6 pt] (0, 0.35)  -- (0, -0.35);
	\draw [<-] (0, -1.25) node [below] {$C'$} -- (0, -0.35);

	
	\begin{scope}[->, dotted, rounded corners]
	
		\draw (0.15, 1.25) -- (0.15, 0.9) -- (0.75, 0.9);
		\draw (-0.5, 0.6) -- (-0.15, 0.6) -- (-0.15, 0.15);
		
		\draw (-0.15, -0.15) -- (-0.15, -0.6) -- (-0.5, -0.6);
		\draw (0.75, -0.9) -- (0.15, -0.9) -- (0.15, -1.25);
	
	\end{scope}

\\
};
\end{tikzpicture}
\caption{\label{3-vert-case-2}An Eulerian circuit with the connections $A \to C, B \to B',$ and $A' \to C'$. For the 3-vertex subgraph, there are two possible ways for the circuit to pass through the internal 3-cycle.}
\end{figure}
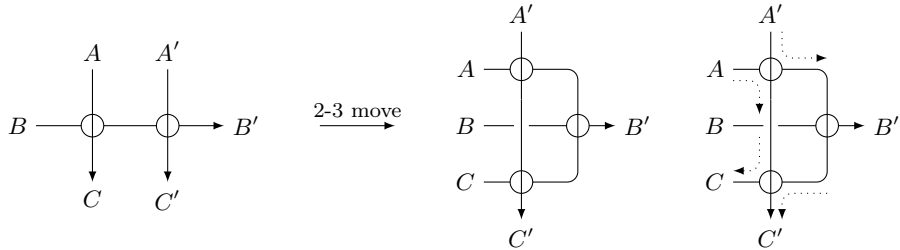

For the 2-vertex subgraph, the sole limitation is there only being one internal edge. On the other hand, the 3-vertex subgraph resulting from the 2-3 move has three internal edges, forming a 3-cycle. This has differing effects on the number of possible connections through the subgraph, and will lead to additional possible ways for the Eulerian circuit to pass through the 3-vertex subgraph. For example, consider the case shown in Figure \ref{3-vert-case-1}, where the connections are $A \to C', B \to C,$ and $A' \to B'$. This choice has a single way of passing through the 2-vertex subgraph. For the 3-vertex subgraph, none of the connected external edges are incident to the same vertex. Because of this, each of the three connections must pass over one internal edge, and so there is also only one possible way for the Eulerian circuit to pass through the 3-vertex subgraph as well. Figure \ref{3-vert-case-2}, by contrast, has the connections $A \to C, B \to B',$ and $A' \to C'$. Thus, there is one connection ($B \to B'$) where both edges are incident to the same vertex; this is the rightmost vertex shown for the 3-vertex subgraph. This gives rise to two possible ways for the connections to occur, depending on which of the other two connections uses two internal edges. Thus, depending on how the Eulerian circuit connects the six external edges, there may be a greater number of possibilities after the 2-3 move than there were before. However, all of the possibilities discussed so far will have induced knot diagrams that are equivalent under the Reidemeister moves.

\begin{figure}[hbt]
\centering
\begin{tikzpicture}[> = latex]
\matrix[column sep = 0.5 cm, row sep = 0.5 cm]{


	\draw [->] (-1.25, 0) node [left] {$B$} -- (1.25, 0) node [right] {$B'$};

	\draw [fill = white] (-0.5, 0) circle (0.15);
	\draw [fill = white] (0.5, 0) circle (0.15);

	\draw [->] (-0.5, 0.75) node [above] {$A$} -- (-0.5, -0.75) node [below] {$C$};
	\draw [->] (0.5, 0.75) node [above] {$A'$} -- (0.5, -0.75) node [below] {$C'$};

	
	\begin{scope}[->, dotted, rounded corners]
	
		\draw (-0.35, 0.625) -- (-0.35, 0.15) -- (0, 0.15);
		\draw (-1, -0.15) -- (-0.65, -0.15) -- (-0.65, -0.75);
		
		\draw (0, -0.15) -- (0.35, -0.15) -- (0.35, -0.75);
		\draw (0.65, 0.625) -- (0.65, 0.15) -- (1, 0.15);
	
	\end{scope}

&

	\draw [->, font = \footnotesize] (0, 0) -- node [above] {2-3 move} (1, 0);

&


	\draw [->] (-0.5, 0) node [left] {$B$} -- (1.5, 0) node [right] {$B'$};
	\draw [fill = white] (0.75, 0) circle (0.15);

	\draw [rounded corners] (-0.5, 0.75) node [left] {$A$} -- (0.75, 0.75) -- (0.75, -0.75) -- (-0.5, -0.75) node [left] {$C$};

	\draw [fill = white] (0, 0.75) circle (0.15);
	\draw [fill = white] (0, -0.75) circle (0.15);

	\draw (0, 1.25) node [above] {$A'$} -- (0, 0.35);
	\draw [draw = white, double = black, double distance between line centers = 3 pt, line width = 2.6 pt] (0, 0.35)  -- (0, -0.35);
	\draw [<-] (0, -1.25) node [below] {$C'$} -- (0, -0.35);

	
	\begin{scope}[->, dotted, rounded corners]
	
		\draw (0.15, 1.25) -- (0.15, 0.9) -- (0.75, 0.9);
		\draw (-0.5, 0.6) -- (-0.15, 0.6) -- (-0.15, 0.15);
		
		\draw (0.15, -0.15) -- (0.6, -0.15) -- (0.6, -0.6);
		\draw (0.9, 0.6) -- (0.9, 0.15) -- (1.5, 0.15);
	
	\end{scope}

\\
};
\end{tikzpicture}
\caption{\label{3-vert-case-1}An Eulerian circuit with the connections $A \to C', B \to C,$ and $A' \to B'$. For the 3-vertex subgraph, there is only one possible way for the circuit to pass through the internal 3-cycle.}
\end{figure}
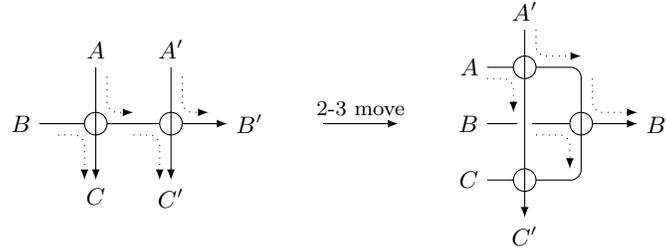

\begin{figure}[hbt]
\centering
\begin{tikzpicture}[> = latex]
\matrix[column sep = 0.5 cm]{


	\draw [->] (-0.5, 0) node [left] {$B$} -- (1.25, 0) node [right] {$B'$};
	\draw [fill = white] (0.75, 0) circle (0.15);

	\draw [rounded corners] (-0.5, 0.75) node [left] {$A$} -- (0.75, 0.75) -- (0.75, -0.75) -- (-0.5, -0.75) node [left] {$C$};

	\draw [fill = white] (0, 0.75) circle (0.15);
	\draw [fill = white] (0, -0.75) circle (0.15);

	\draw [<-] (0, 1.25) node [above] {$A'$} -- (0, 0.35);
	\draw [draw = white, double = black, double distance between line centers = 3 pt, line width = 2.6 pt] (0, 0.35)  -- (0, -0.35);
	\draw [<-] (0, -1.25) node [below] {$C'$} -- (0, -0.35);

	
	\begin{scope}[->, dotted, rounded corners]
	
		\draw (-0.5, -0.6) -- (-0.15, -0.6) -- (-0.15, -0.15);
		\draw (0.6, -0.9) -- (0.15, -0.9) -- (0.15, -1.25);
	
	\end{scope}

&


	\draw [->] (-0.5, 0) node [left] {$B$} -- (1.25, 0) node [right] {$B'$};
	\draw [fill = white] (0.75, 0) circle (0.15);

	\draw [rounded corners] (-0.5, 0.75) node [left] {$A$} -- (0.75, 0.75) -- (0.75, -0.75) -- (-0.5, -0.75) node [left] {$C$};

	\draw [fill = white] (0, 0.75) circle (0.15);
	\draw [fill = white] (0, -0.75) circle (0.15);

	\draw [<-] (0, 1.25) node [above] {$A'$} -- (0, 0.35);
	\draw [draw = white, double = black, double distance between line centers = 3 pt, line width = 2.6 pt] (0, 0.35)  -- (0, -0.35);
	\draw [<-] (0, -1.25) node [below] {$C'$} -- (0, -0.35);

	
	\begin{scope}[->, dotted, rounded corners]
	
		\draw (-0.5, 0.6) -- (-0.15, 0.6) -- (-0.15, 0.15);
		\draw (0.6, 0.9) -- (0.15, 0.9) -- (0.15, 1.25);
	
	\end{scope}

\\
};
\end{tikzpicture}
\caption{\label{3-vert-case-3}An Eulerian circuit with the connections $A \to C', B \to B',$ and $C \to A'$. This set of connections is only realizable in the 3-vertex subgraph.}
\end{figure}
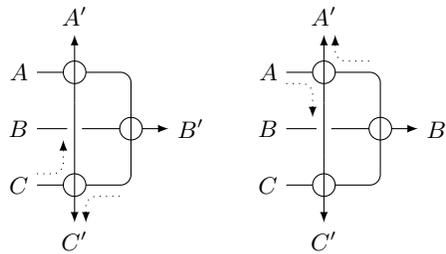

Finally, there are six remaining ways to connect the six external edges, which are not possible with the 2-vertex subgraph, but are with the 3-vertex subgraph. These six connections all involve each external edge with an unprimed label matched with one with a primed label. The number of these extra possibilities depends on how the unprimed and primed labels are matched up, and thus how many edges in the internal 3-cycle are required by each of the three connections. Of the six extra possibilities, up to orientation reversals for each of the connections, two have only one way for the Eulerian circuit to pass through the subgraph, three have two ways (an example is shown in Figure \ref{3-vert-case-3}), and one has six ways. Unlike the previous cases, these new outcomes can have induced knot diagrams that are different than the original 2-vertex subgraph. For 2-vertex subgraphs, at most there could be two crossings: the $B \to B'$ edges passing under both the $A \to A'$ and $C \to C'$ edges. Compare this to Figure \ref{3-vert-case-3}. Although the $B \to B'$ edge still passes under the other two, there are now crossings between the other two edges as well. In the left-hand subgraph, the edge $A \to C'$ passes under the $C \to A'$ edge, while in the right-hand subgraph, it is the reverse. Because of these additional options for the number and type of crossings in the 3-vertex subgraph, and its induced knot diagram, there are additional possibilities for the number of quandle colorings of the graph resulting from the 2-3 move.

\begin{figure}[hbt]
\centering
\begin{tikzpicture}[> = latex]
\matrix[column sep = 0.5 cm]{


	\draw [fill = white] (-1.5, 0) circle (0.15);
	\draw [fill = white] (-0.5, 0) circle (0.15);


	\draw [rounded corners] (-1.65, 0) -- (-2, 0) -- (-2, 0.5) -- (-1.5, 0.5) -- (-1.5, -0.5) -- (0.5, -0.5) -- (0.5, -0.15)
		(0.5, 0.15) -- (0.5, 0.3) -- (0, 0.3) -- (0, -0.3) -- (-0.5, -0.3) -- (-0.5, 0.5) -- (1, 0.5) -- (1, 0) -- (0.15, 0)
		(-1.35, 0) -- (-0.65, 0) (-0.35, 0) -- (-0.15, 0);


	\node at (-0.25, -1.75) {$1^\ell 5^- 7^u 3^-$};

&

	\draw [->, font = \footnotesize] (0, 0) -- node [above] {2-3 move} (1, 0);

&


	\draw [rounded corners] (1.75, -0.15) -- (1.75, -1) -- (-0.5, -1) -- (-0.5, -0.5) -- (-0.15, -0.5)
		(0.15, 0) -- (1.1, 0);


	\draw [fill = white] (0, 0.5) circle (0.15);
	\draw [fill = white] (0.75, 0) circle (0.15);
	\draw [fill = white] (0, -0.5) circle (0.15);


	\draw [rounded corners] (1.4, 0) -- (2.25, 0) -- (2.25, 0.8) -- (0, 0.8) -- (0, -0.8) -- (1.25, -0.8) -- (1.25, 0.3) -- (1.75, 0.3) -- (1.75, 0.15)
		(0.15, 0.5) -- (0.75, 0.5) -- (0.75, -0.5) -- (0.15, -0.5)
		(-0.15, 0.5) -- (-0.5, 0.5) -- (-0.5, 0) -- (-0.15, 0);	


	\node at (0.75, -1.75) {$3^\ell 7^+ 9^+ 1^+ 11^\ell 5^\ell$};

&


	\draw [gray, very thick, rounded corners] (-0.15, 0) -- (-0.5, 0) -- (-0.5, 0.5) -- (-0.15, 0.5)
		(0, 0.65) -- (0, 0.8) -- (2.25, 0.8) -- (2.25, 0) -- (1.4, 0)
		(1.1, 0) -- (0.9, 0)
		(0.75, 0.15) -- (0.75, 0.5) -- (0.15, 0.5)
		(0, 0.35) -- (0, -0.35)
		(-0.15, -0.5) -- (-0.5, -0.5) -- (-0.5, -1) -- (1.75, -1) -- (1.75, -0.15)
		(1.75, 0.15) -- (1.75, 0.3) -- (1.25, 0.3) -- (1.25, -0.8) -- (0, -0.8) -- (0, -0.65)
		(0.15, -0.5) -- (0.75, -0.5) -- (0.75, -0.15)
		(0.6, 0) -- (0.15, 0);

	\draw [gray, very thick] (-0.15, 0.5) arc (180 : 90 : 0.15)
		(0.9, 0) arc (0 : 90 : 0.15)
		(0.15, 0.5) arc (360 : 270 : 0.15)
		(0, -0.35) arc (90 : 180 : 0.15)
		(0, -0.65) arc (270 : 360 : 0.15)
		(0.75, -0.15) arc (270 : 180 : 0.15);
	
\\
};
\end{tikzpicture}
\caption{\label{2-3-ex}The graphs $1^\ell 5^u 7^- 3^-$ and $3^\ell 7^+ 9^+ 1^+ 11^\ell 5^\ell$. To the right is a possible Eulerian circuit through the resulting graph, whose induced knot diagram corresponds to a trefoil knot.}
\end{figure}
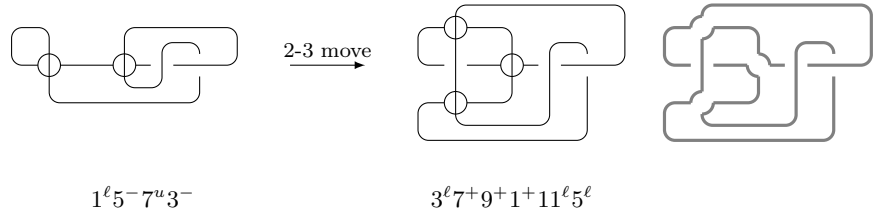

We now give an example of how the quandle coloring invariant polynomial $\Psi_Q$ changes under the 2-3 move. Again, using the Takasaki kei $\mathbb{Z}_3$ (shown in Table \ref{kei}) is sufficient for our purposes. Suppose we consider the 2-3 move acting as in Figure \ref{2-3-ex}. For the graph $1^\ell 5^- 7^u 3^-$ on the left, there are eight possible Eulerian circuits: two choices for the orientation of one of the middle edges (which fixes the other), and two choices each for the orientations of the self-loops on each vertex. The induced knot diagrams for all of these circuits are equivalent to the unknot. This is easily seen for both vertices, as follows. For the left, if the induced knot diagram gives a crossing, this crossing can be eliminated by a RI move; similarly, if there is a crossing for the right vertex, this is eliminated by a RII move. All other crossings in the induced knot diagrams are removed in a similar manner. Therefore, the quandle counting invariant polynomial for the left-hand graph is
\[
	\Psi_{\mathbb{Z}_3} (1^\ell 5^- 7^u 3^-) = 8 u^3
\]
After the 2-3 move, the graph on the right-hand side of Figure \ref{2-3-ex} is obtained, and this graph has 24 possible Eulerian circuits. Changing the structure of the vertices allows additional paths for the circuit to pass through the graph. In particular, there are two Eulerian circuits (ignoring global orientation) where the induced knot diagram has multiple crossings. These crossings result in the appearance the trefoil knot, much like the example for the 1-4 move. One such knot diagram is shown on the right-hand side of Figure \ref{2-3-ex}. As with the previous example in Section \ref{1-4-sect}, this changes the quandle counting invariant polynomial with the addition of a second term, or
\[
	\Psi_{\mathbb{Z}_3} (3^\ell 7^+ 9^+ 1^+ 11^\ell 5^\ell) = 20u^3 + 4u^9
\]
Thus, for both the 1-4 move and the 2-3 move, the Pachner move has added to the set of possible quandle coloring numbers of the graph. Before each move, the graph had three possible colorings of all induced knot diagrams; afterward, the graphs had either three or nine, depending on the Eulerian circuit.

\section{Discussion}
\label{discuss}

The results of this paper provide methods to enumerate and classify knotted 4-regular pseudographs, opening up techniques to answer many interesting questions. Before getting into some of these questions, it is worthwhile to briefly dwell on the enumeration of these graphs. Here, we have developed a nomenclature for knotted 4-regular graphs, based on the Dowker-Thistletwaite sequences of knot theory. Dowker and Thistletwaite also developed an algorithm to enumerate knots~\cite{Dow-Thi83, Hos-Thi-Wee96}, which is easily adapted to knotted graphs as well. This algorithm provides a way of finding all knot diagrams with a certain number of crossings; using knot equivalence moves then finds those knots that are not simply a more complicated diagram for a knot with a lower minimum crossing number. Adapting this algorithm to knotted graphs is straightforward, with the only addition being that nodes can now be either crossings or vertices. This modified algorithm would then obtain all graphs with a given number of nodes. Note that this means we cannot come up with a definite list of all graphs with a certain number of {\it vertices}, since in general, the number of crossings is unlimited. Thus, there is no possibility of having a complete list of all graphs with a given vertex number. With that caveat in mind, we now have a toolbox for building up a census of these graphs, and studying their properties. In both knot theory and abstract graph theory, such a census is crucial in systematically investigating interesting properties.

Now we turn to the questions this census can help answer. Since developing this census hinges crucially on the methods used to characterize and differentiate graphs, exploring these techniques more is an obvious first step. It has been conjectured that there are a finite number of quandles needed to differentiate all knots\cite{Cla-Elh-Sai-Yea14, Dio-Lop03, Fen-Rou92}, and the same question can be asked for knotted graphs as well. Note that, at this point, it is not clear which version would imply the other. Presumably, the number of graphs, for a given number of nodes, would be much greater than the corresponding number of knots with the same number of crossings. This is because all nodes that are crossings can be changed to a vertex state giving at least one induced knot diagram with the same crossing at the vertex. However, this does not account for the use of graph equivalence moves, which may swing the comparative numbers of graphs versus knots in the other direction.

Related to this issue is the choice of invariants used. A relatively simple polynomial invariant based on quandles was used in this paper, but is certainly not the only choice. There are many enhancements of such quandle invariants for knots, and applying them to knotted graphs would be relatively straightfoward. A sampling of these include quandle cohomology~\cite{Car-Jel-Kam-Lan-Sai03}, using quiver structures~\cite{Cho-Nel19}, and the fundamental quandle of a knotted graph~\cite{Nie10}. In particular, quandle cohomology may be an interesting method, especially due to the importance of the Pachner moves in studying states in quantum geometry (such as spin foams). Recall that the Pachner moves on a manifold triangulation are those moves that preserve topological information, so there may be interesting links between cohomology theory on piecewise-linear manifolds and their dual graphs. Examining quantum invariants for graphs -- as mentioned at the end of Section \ref{skein-poly} -- may lead to similar insights from topological quantum field theory.

Continuing with the Pachner moves, as mentioned in Section \ref{Pachner}, there is a partial order on the space of graphs provided by such moves. Using a 1-4 or 2-3 move on a graph increases the number of Eulerian circuits on the final graph. In addition, such moves will generically lead to additional numbers of quandle colorings for the final graph, adding additional terms to the quandle counting invariant polynomial $\Psi_Q$. However, this leaves much to explore. In particular, there is the obvious question if there is a graph invariant that completely characterizes the Pachner partial order. Recall that we defined this ordering $\prec$ earlier such that $G_1 \prec G_2$ if $G_2$ is obtained from $G_1$ by a finite sequence of 1-4 and 2-3 moves. Now, suppose we choose any two distinct graphs $G_a$ and $G_b$, and we wish to see if $G_a \prec G_b$, $G_b \prec G_a$, or else the two graphs are incomparable. Is there a graph invariant that can answer this question for all such graph pairs? It is possible that there is a single type of invariant that provides the answer, but that it must be used with, e.g. a finite set of quandles. Note this is a different question than the more general one of whether there is a finite sequence of moves -- including not only the 1-4 and 2-3 moves, but also their inverses, the 4-1 and 3-2 moves -- relating the two graphs. With this last question, there is some information in the graph itself. The abstract form of the knotted 4-regular graph can be dual to the triangulation of a manifold, and the Pachner moves preserve which manifold the graphs are dual to. Thus, if the two graphs $G_a$ and $G_b$ are dual to different manifolds, then there is no sequence of equivalence moves. The question is whether invariants, such as those discussed in this paper, can provide more information. In particular, it is likely that there exist many pairs of distinct knotted graphs, which are dual to the same manifold when considered as abstract graphs, but their embeddings may prevent the requisite Pachner moves, due to how the edges are placed.

\section{Summary}
\label{conclude}

Knotted graphs appear as components of the quantum states of loop quantum gravity. Thus, it is important to have the ability to enumerate and classify all of these possible graphs. In this paper, we looked at the specific case of 4-regular graphs, and demonstrated two ways of identifying these graphs. The first method, similar to knot polynomials such as the Jones and HOMFLY-PT polynomials, found a polynomial in two abstract variables for every knotted 4-regular graph, with edges oriented using an Eulerian circuit. To ensure that this is independent of the choice of the circuit through the graph diagram, the multiset of all polynomials is obtained. Because this multiset does not change under generalized Reidemeister moves acting on the graph diagram, the multiset is an invariant of the graph. Next, we used the method of quandles to find a second invariant polynomial for the graph. Here, there is a single polynomial of the graph, rather than a multiset, as in the first technique. The quandle counting polynomial is found as the generating function of the multiset listing the number of quandle colorings of the graph diagram, for all Eulerian circuits through the diagram. This polynomial is then independent of the choice of the graph diagram, and thus is also a graph invariant. Using this latter invariant polynomial, the action of the 1-4 and 2-3 Pachner moves was examined. Such moves generically increase the size of the multiset of quandle colorings of the graph diagram, giving a possible method to characterize a partial ordering of all knotted 4-regular graphs under the Pachner moves. These techniques, or enhancements of them, provide a classification method for the space of graphs.



\begin{thebibliography}{99}

	\bibitem{Bor-Maj-Smo96} Borissov, R., Major, S., and Smolin, L., ``The geometry of quantum spin networks," Class. Quant. Grav. 13, 3183--3196 (1996).
	\bibitem{Bur} Burton, B. A., ``Face pairing graphs and 3-manifold enumeration," J. Knot Theory Ramifications 13, 1057--1101(2004); Burton, B. A., ``Detecting genus in vertex links for the fast enumeration of 3-manifold triangulations," ISSAC 2011: Proceedings of the 36th International Symposium on Symbolic and Algebraic Computation, 59--66 (2011).
	\bibitem{Car-Jel-Kam-Lan-Sai03} Carter, J.S, Jelsovsky, D., Kamada, S., Langford, L., and Saito, M., ``Quandle cohomology and state-sum invariants of knotted curves and surfaces," Trans. of Amer. Math. Soc. 355, 3947--3989 (2003).
	\bibitem{Cho-Nel19} Cho, K. and Nelson, S., ``Quandle coloring quivers," J. Knot Theory Ramifications 28, 1950001 (2019) 
	\bibitem{Cla-Elh-Sai-Yea14} Clark, W.E., Elhamdadi, M., Saito, M., and Yeatman, T., ``Quandle colorings of knots and applications," J. Knot Theory Ramifications 23, 1450035 (2014). 
	\bibitem{DeP-Rov96} De Pietri, R. and Rovelli, C., ``Geometry eigenvalues and scalar product from recoupling theory in loop quantum gravity," Phys. Rev. D54, 2664--2690 (1996). 
	\bibitem{Dio-Lop03} Dionisio, F. M. and Lopes, P., ``Quandles at finite temperatures II," J. Knot Theo. Ramifications 12, 1041--1092 (2003). 
	\bibitem{Dit-Ste14} Dittrich, B. and Steinhaus, S., ``Time evolution as refining, coarse graining and entangling," New J. Phys. 16, 123041 (2014). 
	\bibitem{Dow-Thi83} Dowker, C. H., and Thistlethwaite, M. B., ``Classification of knot projections," Topology Appl. 16, 19--31 (1983).
	\bibitem{Elh-Nel15} Elhamdadi, M. and Nelson, S., {\it Quandles: An introduction to the algebra of knots} (American Mathematical Society, Providence, RI, 2015).
	\bibitem{Fai-Rov04} Fairbairn, W., and Rovelli, C., ``Separable Hilbert space in loop quantum gravity," J. Math. Phys. 45, 2802--2814 (2004). 
	\bibitem{Fen-Rou92} Fenn, R. and Rourke, C., ``Racks and links in codimension two," J. Knot Theo. Ramifications 01, 343--406 (1992).
	\bibitem{Gro-Rov96} Grot, N. and Rovelli, C., ``Moduli-space structure of knots with intersections," J. Math. Phys. 37, 3014--3021 (1996). 
	\bibitem{Hos-Thi-Wee96} Hoste, J., Thistlethwaite, M., and Weeks, J., ``The first 1,701,936 knots," Math. Intelligencer 20, 33 -- 48 (1996).
	\bibitem{Kau89} Kauffman, L. H., ``Invariants of graphs in three-space," Trans. Amer. Math. Soc. 311, 697--710 (1989).
	\bibitem{Nie10} Niebrzydowski, M., ``Coloring invariants of spatial graphs," J. Knot Theory Ramifications 19, 829--841 (2010). 
	\bibitem{Pac91} Pachner, U., ``P.L. homeomorphic manifolds are equivalent by elementary shellings," Eur. J. Combin. 12, 129--145 (1991).
	\bibitem{Smo-Wan08} Smolin, L. and Wan, Y., ``Propagation and interaction of chiral states in quantum gravity," Nucl. Phys. B796, 331-359 (2008).
	\bibitem{Ste20} Steinhaus, S., ``Coarse graining spin foam quantum gravity - a review," Frontiers in Physics 8, 295 (2020).
	\bibitem{Wan07} Wan, Y., ``On braid excitations in quantum gravity," arXiv:0710.1312v1.

\end{thebibliography}
\end{document}